\documentclass[journal]{IEEEtran}

\usepackage{epsfig,graphics}
\usepackage{amsmath,amssymb,amsthm,multirow}
\usepackage{cite}
\usepackage{subfigure}
\usepackage{balance}

\newtheorem{theorem}{Theorem}
\newtheorem{Property}{Property}

\newtheorem{assumption}{Assumption}

\usepackage{color}  

\newcommand{\cb}[1]{{\boldsymbol{#1}}}
\newcommand{\cp}[1]{\ifmmode {\mathcal{#1}}\else ${\mathcal{#1}}$\fi}
\newcommand{\cbp}[1]{{\boldsymbol{\mathcal{#1}}}}

\newcommand{\bpsi}{\boldsymbol{\psi}}

\newcommand{\bpsit}{\widetilde\bpsi}

\newcommand{\bu}{\boldsymbol{u}}
\newcommand{\bp}{\boldsymbol{p}}

\newcommand{\bg}{\boldsymbol{g}}
\newcommand{\bw}{\boldsymbol{w}}
\newcommand{\bx}{\boldsymbol{x}}
\newcommand{\by}{\boldsymbol{y}}

\newcommand{\bwt}{\widetilde\bw}

\newcommand{\bA}{\boldsymbol{A}}
\newcommand{\bB}{\boldsymbol{B}}
\newcommand{\bC}{\boldsymbol{C}}
\newcommand{\bD}{\boldsymbol{D}}
\newcommand{\bQ}{\boldsymbol{Q}}

\newcommand{\bP}{\boldsymbol{P}}

\newcommand{\br}{\boldsymbol{r}}

\newcommand{\bR}{\boldsymbol{R}}

\newcommand{\bX}{\boldsymbol{X}}

\newcommand{\bM}{\boldsymbol{M}}

\newcommand{\bSig}{\boldsymbol{\Sigma}}
\newcommand{\bsig}{\boldsymbol{\sigma}}

\newcommand{\bGam}{\boldsymbol{\Gamma}}
\newcommand{\bI}{\boldsymbol{I}}

\newcommand{\expec}[1]{\mathbb{E}\{#1\}}
\newcommand{\meanM}{\overline{\boldsymbol{M}}}
\newcommand{\meanA}{\overline{\bA}}
\newcommand{\meanP}{\overline{\bP}}

\newcommand{\N}{{\cp{N}}}
\newcommand{\C}{{\cp{C}}}

\newcommand{\tr}{\text{trace}}

\newcommand{\bvc}{\text{bvec}}
\newcommand{\col}{\text{col}}

\begin{document}

\title{Multitask diffusion adaptation over asynchronous networks}
\author{Roula Nassif, \IEEEmembership{Student Member, IEEE}, C{\'e}dric Richard, \IEEEmembership{Senior Member, IEEE}\\
Andr{\'e} Ferrari, \IEEEmembership{Member, IEEE}, Ali H. Sayed, \IEEEmembership{Fellow Member, IEEE}
\thanks{The work of C. Richard and A. Ferrari was partly supported by ANR and DGA grant ANR-13-ASTR-0030 (ODISSEE project). The work of A. H. Sayed was supported in part by NSF grants CIF-1524250 and ECCS-1407712. A short version of this work appears in the conference publication~\cite{Nassif2014asilomar}.

R. Nassif, C. Richard, and A. Ferrari are with the Universit{\'e} de Nice Sophia-Antipolis, France (email: roula.nassif@oca.eu; cedric.richard@unice.fr; andre.ferrari@unice.fr).

A. H. Sayed is with the department of electrical engineering, University of California, Los Angeles, USA (email: sayed@ee.ucla.edu). }
}

\maketitle

%
%

\begin{abstract}
The multitask diffusion LMS is an efficient strategy to simultaneously infer, in a collaborative manner, multiple parameter vectors. Existing works on multitask problems assume that all agents respond to data synchronously. In several applications, agents may not be able to act synchronously because networks can be subject to several sources of uncertainties such as changing topology, random link failures, or agents turning on and off for energy conservation. In this work, we describe a model for the solution of multitask problems over asynchronous networks and carry out a detailed mean and mean-square error analysis. Results show that sufficiently small step-sizes can still ensure both stability and performance. Simulations and illustrative examples are provided to verify the theoretical findings. 
\end{abstract}


\section{Introduction}

Distributed adaptive learning enables agents to learn a concept via local information exchange, and to continuously adapt to track possible concept drifts. Distributed implementations offer an attractive alternative to centralized solutions with advantages related to scalability, robustness, and decentralization (see, e.g.,~\cite{Sayed2014Proc,sayed2014adaptation} and the many examples therein). Several strategies for distributed online parameter estimation have been 
proposed in the literature, including consensus strategies~\cite{Tsitsiklis1984,Xiao2004,Braca2008,Nedic2009,Kar2009,Srivastava2011}, incremental strategies~\cite{Bertsekas1997,Nedic2001,Rabbat2005,Blatt2007,Lopes2007incr}, and diffusion strategies~\cite{Sayed2013diff,Sayed2013intr,Lopes2008diff,Cattivelli2010diff,ChenUCLA2012,ChenUCLA2013}. Incremental techniques operate on a cyclic path that runs across all nodes, which makes them sensitive to link failures and problematic for adaptive implementations. On the other hand, diffusion strategies are particularly attractive due to their enhanced adaptation performance and wider stability ranges than consensus-based implementations. Accessible overviews of results on diffusion adaptation can be found in~\cite{Sayed2013diff,Sayed2013intr,Sayed2014Proc}. 

Most prior literature focuses primarily on the case where nodes estimate a single parameter vector collaboratively. We refer to problems of this type as single-task problems. Some applications require more complex models and flexible algorithms than single-task implementations since their agents may involve the need to track multiple targets simultaneously. For instance, sensor networks deployed to estimate a spatially-varying temperature profile need to exploit more directly the spatio-temporal correlations that exist between measurements at neighboring nodes~\cite{Abdolee2014}. Likewise, monitoring applications where agents need to track the movement of multiple correlated targets need to exploit the correlation profile in the data for enhanced accuracy. Problems of this kind, where nodes need to infer multiple parameter vectors, are referred to as multitask problems. 

Existing strategies to address multitask problems mostly depend on how the tasks relate to each other and on exploiting some prior information. There have been some useful works dealing with such problems over distributed networks. For example, in~\cite{Bogdanovic2014} a diffusion strategy of the LMS type is developed to solve distributed optimization problems where nodes are interested in estimating parameters of local interest and parameters of global interest to the whole network. In~\cite{plata2014distributed}, an extension of the diffusion algorithm developed in~\cite{Bogdanovic2014} allows nodes to estimate parameters of common interest to a subset of nodes simultaneously with parameters of local and global interest. In comparison, the parameter space is decomposed into two orthogonal subspaces in~\cite{chen2014diffusion3}, with one of the subspaces being common to all nodes. Multitask estimation algorithms over fully connected networks and tree networks are also considered in~\cite{Bertrand2010P1,Bertrand2011}. These works assume that the node-specific parameter vectors lie in a common latent signal subspace and exploit this property to compress information and to reduce communication costs. An alternative way to exploit and model relationships among tasks is to formulate optimization problems with appropriate co-regularizers between nodes~\cite{chen2013multitask,nassif2015icassp}. The multitask diffusion LMS algorithm derived in~\cite{chen2013multitask} relies on this principle, and we build on this construction in this article. In this context, the network is not assumed to be fully connected and agents need not be interested in some common parameters. It is sufficient to assume that different clusters within the network are interested in their own models, and that there are some correlations among the models of adjacent clusters. These correlations are captured by means of regularization parameters. Multitask estimation problems have also been addressed over diffusion networks where no prior information on possible relationships between tasks is assumed and nodes do not know which other nodes share the same task~\cite{chen2013pareto,zhao2012clustering,chen2015diffusioovermultitask,nassif2015noisylinks}. In this case, it was argued in~\cite{chen2013pareto} that the diffusion iterates converge to a Pareto optimal solution when confronted with a multi-objective optimization problem. To avoid cooperation between neighbors seeking different objectives, automatic clustering techniques using diffusion strategies have been proposed. The clustering techniques developed in~\cite{zhao2012clustering,chen2015diffusioovermultitask} are based on setting the combination coefficients in an online manner. The technique proposed in~\cite{zhao2015clustering} is based on solving a hypothesis test problem for setting the neighborhood in an online manner.

The aforementioned works on multitask problems assume that all agents respond to data synchronously. In several applications, agents may not be able to act synchronously because networks can be subject to several sources of uncertainties such as changing topology, random link failures, or agents turning on and off. There exist several useful studies in the literature on the performance of consensus and gossip strategies in the presence of asynchronous events~\cite{Tsitsiklis1986,Boyd2006,Kar2009,Srivastava2011} or changing topologies \cite{Kar2011,Boyd2006,Kar2009,Srivastava2011,Kar2008,Aysal2009,Kar2010,Jakovetic2010,Jakovetic2011}. In most parts, these works investigate pure averaging algorithms that cannot process streaming data or the works assume noise-free data or make use of decreasing step-size sequences. There are also studies in the context of diffusion strategies. In particular, the works~\cite{SayedPart1,SayedPart2,SayedPart3} advanced a rather general framework for asynchronous networks that includes many prior models as special cases. The works examined how asynchronous events interfere with the behavior of adaptive networks in the presence of streaming noisy data and under constant step-size adaptation. Several interesting conclusions are reported in~\cite{SayedPart3} where comparisons are carried out between synchronous and asynchronous behavior, as well as with centralized solutions. In the current work, we would like to examine similar effects to \cite{SayedPart1},\cite{SayedPart2} albeit in the context of multitask networks as opposed to single-task networks. In this case, a new dimension arises in that asynchronous events can interfere with the exchange of information among clusters. We examine in some detail the mean and mean-square stability of the multitask network and show that sufficiently small step-sizes can still ensure convergence and performance. Various simulation results illustrate the theoretical findings.

This paper is organized as follows. In Section~\ref{sec:Multitask Diffusion LMS}, we briefly recall the multitask diffusion LMS strategy and we introduce a fairly general model for asynchronous behavior. Under this model, agents in the network may stop updating their solutions, or may stop sending or receiving information in a random manner. Section~\ref{sec:Stochastic Performance} analyzes the theoretical performance of the algorithm, in the mean and mean-square error sense. In Section~\ref{simulation}, experiments are presented to illustrate the performance of the diffusion multitask approach over asynchronous networks.

\section{Multitask diffusion LMS over asynchronous networks}
\label{sec:Multitask Diffusion LMS}

Before starting our presentation, we provide a summary of some of the main symbols used in the article. Other symbols will be defined in the context where they are used:

\medskip

\begin{tabular}{ll}
\hspace{-0.4cm} $x$ 			&\hspace{-0.55cm}	 Normal font letters denote scalars. \\
\hspace{-0.3cm}$\bx$ 			&\hspace{-0.55cm} Boldface lowercase letters denote column vectors. \\
\hspace{-0.3cm}$\bR$ 			& \hspace{-0.4cm}Boldface uppercase letters denote matrices.\\ 
\hspace{-0.3cm}$(\cdot)^\top$ 		&\hspace{-0.55cm}  Matrix transpose.\\
\hspace{-0.3cm}$(\cdot)^{-1}$ 		& \hspace{-0.55cm} Matrix inverse.\\
\hspace{-0.3cm}$\bI_N$ 			&\hspace{-0.55cm}  Identity matrix of size $N \times N$.\\
\hspace{-0.3cm}$\cp{N}_k$ 		& \hspace{-0.55cm} The set of nodes containing the neighborhood of\\
\hspace{-0.3cm}$ $ 				&\hspace{-0.5cm}	node $k$, including $k$.\\
\hspace{-0.3cm}$\cp{N}_{k}^{-}$ 	& \hspace{-0.55cm} The set of nodes containing the neighborhood of\\
\hspace{-0.3cm}$ $ 				&\hspace{-0.5cm}	node $k$, excluding $k$.\\
\hspace{-0.3cm}$\C_j$			&\hspace{-0.5cm}  Cluster $j$, i.e., index set of nodes in the $j$-th cluster.\\
\hspace{-0.3cm}$\C(k)$			& \hspace{-0.5cm} The cluster of nodes to which node $k$ belongs, \\
\hspace{-0.3cm}$$				&\hspace{-0.45cm} including $k$.\\
\hspace{-0.3cm}$\C(k)^{-}$		& \hspace{-0.5cm} The cluster of nodes to which node $k$ belongs, \\
\hspace{-0.3cm}$$				&\hspace{-0.35cm}excluding $k$.\\
\end{tabular} 

\bigskip

\noindent We now briefly recall the \emph{synchronous} diffusion adaptation strategy developed in \cite{chen2013multitask} for solving distributed optimization problems over multitask networks.

\subsection{Multitask diffusion adaptation}

\begin{figure}[!h]
	\centering
	\includegraphics[trim = 3mm 3mm 3mm 50mm, clip, scale=0.5]{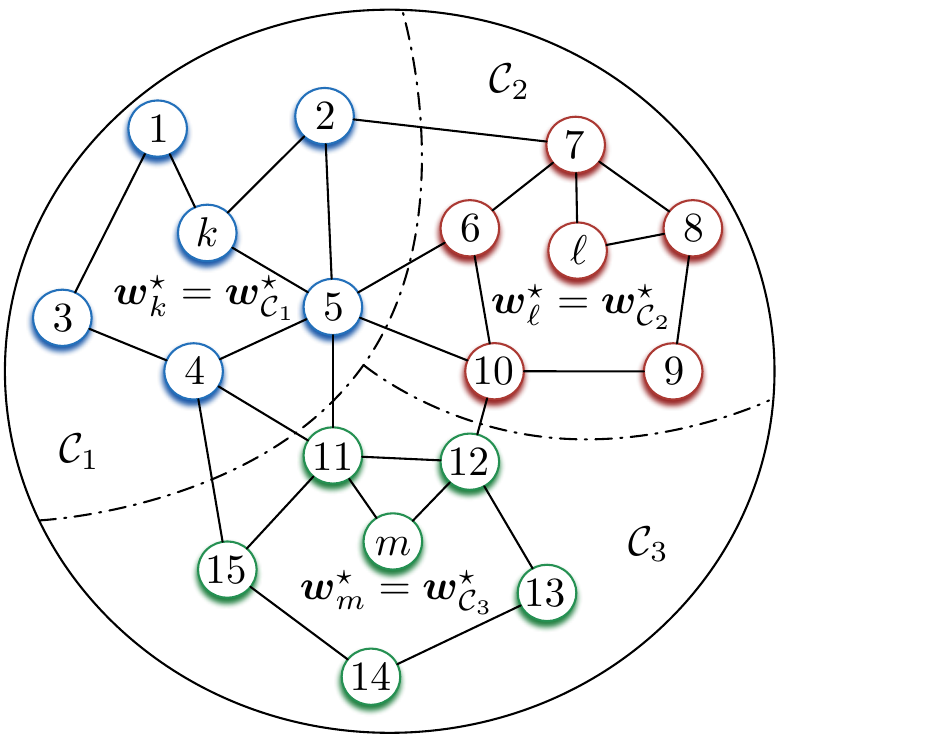}
	\caption{Clustered network consisting of $3$ clusters. Two clusters are connected if they share at least one edge.}
	\label{fig: clustered network}
\end{figure}

We consider a connected network consisting of $N$ nodes grouped into $Q$ clusters, as illustrated in Figure \ref{fig: clustered network}. The problem is to estimate an $L\times 1$ unknown vector $\bw^{\star}_k$ at each node $k$ from collected data. Node $k$ has access to temporal measurement sequences $\{d_k(i), \bx_k(i)\}$, where $d_k(i)$ is a scalar zero-mean reference signal, and $\bx_k(i)$ is an $L\times 1$ regression vector with a positive-definite covariance matrix $ \bR_{x,k}=E\{\bx_k(i)\bx_k^\top(i)\} >0$. The data at node $k$ are assumed to be related via the linear regression model
	\begin{equation}
		\label{eq:datamodel}
		d_k(i)=\bx_k^\top(i)\, \bw_k^\star + z_k(i),
	\end{equation}
where $z_k(i)$ is a zero-mean i.i.d. noise of variance $\sigma_{z,k}^2$ that is independent of any other signal. We assume that nodes belonging to the same cluster have the same parameter vector to estimate, namely,
\begin{equation}
	\label{eq:cluster}
	\bw^{\star}_k=\bw^{\star}_{\C_q},\quad\text{whenever}\quad k\in\C_q.
\end{equation}
We say that two clusters are connected if there exists at least one edge linking a node from one cluster to a node in the other cluster. We also assume that relationships between connected clusters exist so that cooperation among adjacent clusters is beneficial. In particular, we suppose that the parameter vectors corresponding to two connected clusters $\C_p$ and $\C_q$ satisfy certain properties, such as being close to each other  \cite{chen2013multitask}. Cooperation across these clusters can therefore be beneficial to infer $\bw^{\star}_{\C_p}$ and $\bw^{\star}_{\C_q}$.

Consider the cluster $\C(k)$ to which node $k$ belongs. A local cost function, $J_k(\bw_{\C(k)})$, is associated with node $k$. It is assumed to be strongly convex and second-order differentiable, an example of which is the mean-square error criterion considered throughout this paper and defined by
\begin{equation}
	\label{eq:JkMSE}
	J_k(\bw_{\C(k)}) = \expec{|d_k(i) - \bx_k^\top(i)\,\bw_{\C(k)}|^2}.
\end{equation}
Depending on the application, there may be certain properties among the optimal vectors $\{\bw_{\C_1}^{\star},\ldots,\bw_{\C_Q}^{\star}\}$ that deserve to be promoted in order to enhance estimation accuracy. Among other possible options, a smoothness condition was enforced in~\cite{chen2013multitask}. Specifically, the local variation of the graph signal at node $k$ was defined as the squared $\ell_2$-norm of the graph gradient at this node~\cite{grady2010discrete}, namely,
\begin{equation}
	\label{eq:graphgrad}
	\|\nabla_k\cp{W}\|^2 = \sum_{\ell\in \N_k}\rho_{k\ell}\|\bw_k-\bw_\ell\|^2
\end{equation}
where $\rho_{k\ell}$ is a nonnegative weight assigned to the edge between nodes $k$ and $\ell$. As an alternative to \eqref{eq:graphgrad} and in order to promote piecewise constant transitions in the entries of the parameter vectors, the use of the $\ell_1$-norm of the graph gradient at each node was also proposed and studied in~\cite{nassif2015icassp}.  In this paper, we will focus on \eqref{eq:graphgrad}.

To estimate the unknown parameter vectors $\bw^\star_{\C_1},\ldots,\bw^\star_{\C_Q}$, it was shown in \cite{chen2013multitask} that the local cost~\eqref{eq:JkMSE} and the regularizer~\eqref{eq:graphgrad} can be combined at the level of each cluster. This formulation led to the following estimation problem defined in terms of $Q$ Nash equilibrium problems~\cite{basar1995dynamic}, where each cluster $\C_j$ estimates $\bw^\star_{\C_j}$ by minimizing the regularized cost function ${J_{\C_j}}(\bw_{\C_j},\bw_{-\C_j})$:
\begin{equation}
	\label{eq:P}
	(\cp{P}_j)
	\left\lbrace
	\begin{split}
		&\min_{\bw_{\C_j}}{J_{\C_j}}(\bw_{\C_j},\bw_{-\C_j}) \\
		&\text{with }\; {J_{\C_j}}(\bw_{\C_j},\bw_{-\C_j}) \\
			&\qquad\quad= \sum_{k\in\C_j}  \expec{|d_k(i) - \bx_k^\top(i)\,\bw_{\C(k)}|^2} \\
			&\qquad\quad+ \eta\,\sum_{k\in\C_j}\sum_{\ell\in\N_k\backslash\C_j} \rho_{k\ell}  \,\|\bw_{\C(k)} - \bw_{\C(\ell)}\|^2
	\end{split} \right.
\end{equation}
for $j = 1, \dots, Q$. Note that we have kept the notation $\bw_{\C(k)}$ in \eqref{eq:P} to make the role of the regularization term clearer, even though we have $\bw_{\C(k)}=\bw_{\C_j}$ for all $k$ in $\C_j$. The notation $\bw_{-\C_j}$ denotes the collection of weight vectors estimated by the other clusters, that is, $\bw_{-\C_j} =\{\bw_{\C_q}:q=1,\ldots,Q\}\setminus\{\bw_{\C_j}\}$. The second term on the RHS of expression~\eqref{eq:P} enforces smoothness of the resulting graph parameter vectors $\{\bw_{\C_1},\ldots,\bw_{\C_Q}\}$, with strength parameter $\eta \geq 0$. In~\cite{chen2013multitask}, the coefficients~$\{\rho_{k\ell}\}$ were chosen to satisfy the conditions:
\begin{equation}
	\sum_{\ell\in \N_k\setminus \C(k)^-}\rho_{k\ell}=1, 
	\text{ and }
	\left\lbrace\begin{array}{lr}
	\rho_{k\ell} > 0, \text{ if } \ell\in \N_k\setminus\C(k), \\
	\rho_{kk} \geq 0, \\
	\rho_{k\ell}=0, \text{ otherwise.} 
	\end{array}\right.
\end{equation}
We impose $\rho_{k\ell}=0$ for all $\ell\notin\N_k\setminus\C(k)$ since nodes belonging to the same cluster estimate the same parameter vector.

Following the same line of reasoning from~\cite{Sayed2013intr,Cattivelli2010diff} in the single-task case, and extending the argument to problem~\eqref{eq:P} by using Nash-equilibrium properties \cite{basar1995dynamic,rosen1965existence}, the following diffusion strategy of the adapt-then-combine (ATC) form was derived in \cite{chen2013multitask} for solving the multitask learning problem \eqref{eq:P} in a distributed manner:
\begin{equation}
	\label{synchronous_ATC}
	\left\lbrace
	\begin{split}
	\bpsi_{k}(i+1)&=\bw_{k}(i)+\mu_k\,\bx_k(i)\,(d_k(i)-\bx_k^\top(i)\,\bw_k(i))\\
			     &\quad+\eta\,\mu_k\Big(\hspace{-1mm}\sum_{\ell\in \N_k\setminus\C(k)^-}\hspace{-2mm}\rho_{k\ell}(\bw_{\ell}(i)-\bw_{k}(i))\Big),\\
	\bw_{k}(i+1)&=\hspace{-2mm}\sum_{\ell\in \N_k \cap \C(k)}\hspace{-2mm}a_{\ell k}\,\bpsi_{\ell}(i+1).
	\end{split}\right.
\end{equation}
where $\bw_{k}(i)$ denotes the estimate of the unknown parameter vector $\bw^\star_{k}$ at node $k$ and iteration $i$, and $\mu_k$ is a positive step-size parameter. The combination coefficients $\{a_{\ell k}\}$ are nonnegative scalars that are chosen to satisfy the conditions:
\begin{equation}
	\sum_{\ell\in \N_k\cap \C(k)}a_{\ell k}=1, \text{ and }
	\left\lbrace\begin{array}{lr}
	a_{\ell k} > 0, \text{ if } \ell\in \N_{k}\cap \C(k),\\
	a_{\ell k}=0, \text{ otherwise.}
	\end{array}\right.
\end{equation}
There are several ways to select these coefficients such as using the averaging rule or the Metropolis rule (see~\cite{Sayed2013intr} for a listing of these and other choices).

\subsection{Asynchronous multitask diffusion adaptation}
\label{asynchronous model}

To model the asynchronous behavior over networks, we follow the same procedure developed in \cite{SayedPart1} since the model presented in that work allows us to cover many situations of practical interest. Specifically, we replace each deterministic step-size $\mu_k$ by a random process $\mu_k(i)$, and model uncertainties in the links by using \textit{random} combination coefficients $\{a_{\ell k}(i)\}$ and \textit{random} regularization factors $\{\rho_{k\ell}(i)\}$. In other words, we modify the multitask diffusion strategy \eqref{synchronous_ATC} to the following form:
\begin{equation}
	\label{asynchronous_ATC}
	\left\lbrace
	\begin{split}
	\bpsi_{k}(i+1)&=\bw_{k}(i)+\mu_k(i)\,\bx_k(i)\,(d_k(i)-\bx_k^\top(i)\,\bw_k(i))\\
			    &\quad+\eta\,\mu_k(i)\Big(\hspace{-2mm}\sum_{\ell\in \N_k(i)\setminus\C(k)^-}\hspace{-3.5mm}
		\rho_{k\ell}(i)(\bw_{\ell}(i)-\bw_{k}(i))\Big),\\
	\bw_{k}(i+1)&=\hspace{-2mm}\sum_{\ell\in \N_k(i) \cap \C(k)}\hspace{-3mm}a_{\ell k}(i)\,\bpsi_{\ell}(i+1)
	\end{split}\right.
\end{equation}
where $\N_{k}(i)$ is also now random and denotes the random neighborhood of agent $k$ at time instant $i$. The composition of each cluster is assumed to be known a priori and does not change over time. When dealing with multitask networks, compared to single-task networks~\cite{SayedPart1}, a second source of uncertainty comes from links transmitting data between clusters. Indeed, data transmitted over intra-cluster links are used to reach a consensus while data transmitted over inter-cluster links are used to promote relationships between tasks.  In a manner similar to \cite{SayedPart1}, the asynchronous network model is assumed to satisfy the following conditions:
\begin{itemize}
	\item\emph{Conditions on the step-size parameters:} At each time instant $i$, the step-size at node $k$ is a bounded nonnegative random variable $\mu_k(i)\in[0,\mu_{\max,k}]$. These step-sizes are collected into the random matrix $\bM(i)\triangleq\text{diag}\{\mu_1(i),\ldots,\mu_N(i)\}$. We assume that $\{\bM(i), i\geq 0\}$ is a weakly stationary random process with mean $\meanM$ and Kronecker-covariance matrix $\bC_M$ of size $N^2\times N^2$ defined as
\begin{equation}
	\label{eq:CM}
	\bC_M \triangleq \expec{(\bM(i)-\meanM)\otimes(\bM(i)-\meanM)}
\end{equation}
with $\otimes$ denoting the Kronecker product.

	\item\emph{Conditions on the combination coefficients:} The random coefficients $\{a_{\ell k}(i)\}$ used to scale the estimates $\{\bpsi_\ell(i+1)\}$ that are being received by node $k$ from its cluster neighbors $\ell\in\N_{k}(i)\cap \C(k)$ satisfy the following constraints at each iteration $i$:
\begin{equation}
	\label{eq:Aconstraint}
	\hspace{-0.25mm}\sum_{\ell\in \N_k(i)\cap \C(k)}\hspace{-6.75mm}a_{\ell k}(i)=1, \text{and}
	\left\lbrace\begin{array}{lr}
	\hspace{-2mm}a_{\ell k}(i) > 0, \text{if }\ell\in \N_{k}(i)\cap \C(k)\\
	\hspace{-2mm}a_{\ell k}(i)=0, \text{otherwise.}
	\end{array}\right.
\end{equation}
We collect these coefficients into the random $N\times N$ left-stochastic matrix $\bA(i)$. We again assume that $\{\bA(i), i\geq 0\}$ is a weakly stationary random process. Let $\meanA$ be its mean and $\bC_A$ its Kronecker-covariance matrix of size $N^2\times N^2$ defined as
\begin{equation}
	\bC_A \triangleq \expec{(\bA(i)-\meanA)\otimes(\bA(i)-\meanA)}.
\end{equation}

\item\emph{Conditions on the regularization factors:} The random factors $\{\rho_{k\ell}(i)\}$, which adjust the regularization strength between the parameter vectors at neighboring nodes of distinct clusters, satisfy the following constraints at each iteration $i$:
\begin{equation}
	\label{eq:Pconstraint}
	\hspace{-0.25mm}\sum_{\ell\in \N_k(i)\setminus \C(k)^-}\hspace{-7.25mm}\rho_{k\ell}(i)=1, 
	\text{and}
	\left\lbrace\begin{array}{lr}
	\hspace{-2mm}\rho_{k\ell}(i) > 0, \text{if } \ell\in \N_k(i)\setminus\C(k) \\
	\hspace{-2mm}\rho_{kk}(i) \geq 0, \\
	\hspace{-2mm}\rho_{k\ell}(i)=0, \text{otherwise.} 
	\end{array}\right.
\end{equation}
We collect these coefficients into the random ${N\times N}$ right-stochastic matrix $\bP(i)$. We assume that $\{\bP(i), i\geq 0\}$ is a weakly stationary random process with mean $\meanP$ and Kronecker-covariance matrix $\bC_P$ of size $N^2\times N^2$ defined as
\begin{equation}
	\bC_P \triangleq \expec{(\bP(i)-\meanP)\otimes(\bP(i)-\meanP)}.
\end{equation}
	
	\item\emph{Independence assumptions:} To enable tractable analysis, we shall assume that the random matrices $\bM(i)$, $\bA(i)$, and $\bP(i)$ at iteration $i$ are mutually-independent and independent of any other random variables. These matrices are related to node, intra-cluster and inter-cluster link failures, respectively.

	\item{\textit{Mean graph:}} The mean matrices $\meanA$ and $\meanP$ define the intra-cluster and inter-cluster neighborhoods, namely,  $\N_k\cap\C(k)$ and $\N_k\setminus\C(k)$ for all $k$, respectively. We refer to the neighborhoods $\N_k=\big(\N_k\cap\C(k)\big)\cup\big(\N_k\setminus\C(k)\big)$ for all $k$, defined by $\meanA$ and $\meanP$, as the mean graph. 
	\noindent In view of the above conditions, the mean combination coefficients $\bar{a}_{\ell k}\triangleq\expec{a_{\ell k}(i)}$ and regularization factors $\bar{\rho}_{k\ell }\triangleq\expec{\rho_{k\ell}(i)}$ are nonnegative and satisfy the following constraints.
\begin{equation}
	\sum_{\ell\in \N_k\cap \C(k)}\bar{a}_{\ell k}=1, \text{ and }
	\left\lbrace\begin{array}{lr}
	\bar{a}_{\ell k} > 0, \text{ if } \ell\in \N_{k}\cap \C(k),\\
	\bar{a}_{\ell k}=0, \text{ otherwise,}
	\end{array}\right.
\end{equation}
\begin{equation}
	\sum_{\ell\in \N_k\setminus \C(k)^-}\bar\rho_{k\ell}=1, 
	\text{ and }
	\left\lbrace\begin{array}{lr}
	\bar\rho_{k\ell} > 0, \text{ if } \ell\in \N_k\setminus\C(k), \\
	\bar\rho_{kk} \geq 0, \\
	\bar\rho_{k\ell}=0, \text{ otherwise.} 
	\end{array}\right.
\end{equation}
\end{itemize}	
\medskip
Using the same arguments as Lemmas 2 and 3 in~\cite{SayedPart1}, we can state the following properties for the asynchronous model~\eqref{asynchronous_ATC}.

{\Property The $N \times N$ matrix $\meanA$ and the $N^2 \times N^2$ matrix $\meanA\otimes\meanA+\bC_A$ are left-stochastic matrices.}

{\Property The $N \times N$ matrix $\meanP$ and the $N^2 \times N^2$ matrix $\meanP\otimes\meanP+\bC_P$ are right-stochastic matrices.}

{\Property For every node $k$, the neighborhood $\N_k$ that is defined by the mean graph of the asynchronous model~\eqref{asynchronous_ATC} is equal to the union of all possible realizations for the random neighborhood $\N_k(i)=\big(\N_k(i)\cap\C(k)\big)\cup\big(\N_k(i)\setminus\C(k)\big)$}.

\medskip

We provide in Appendix~\ref{bernoulli model} one example for a common asynchronous network referred to as the Bernoulli network. The Bernoulli model proposed in~\cite{SayedPart1} is more general than the one used for modeling random link failures in consensus networks~\cite{Kar2008,Kar2009} since it also allows to consider random ``on-off" behavior for agents. When dealing with multitask problems over asynchronous network, additional sources of uncertainties must be considered. The network provided in Appendix~\ref{bernoulli model} allows us to jointly model intra-cluster link failures, inter-cluster link failures, and random ``on-off" behaviors for agents.

\section{Performance of multitask diffusion over asynchronous networks}
\label{sec:Stochastic Performance}

The performance of the multitask diffusion algorithm \eqref{asynchronous_ATC} is affected by various random perturbations due to the asynchronous events. We now examine the stochastic behavior of this strategy in the mean and mean-square error sense. 

\subsection{Mean error behavior analysis}
For each agent $k$, we introduce the weight error vectors:
\begin{equation}
	\bwt_k(i)\triangleq\bw_k^\star-\bw_k(i), \quad\quad\quad\bpsit_k(i)\triangleq\bw_k^\star-\bpsi_k(i)
\end{equation}
where $\bw_k^\star$ is the optimum parameter vector at node $k$. We denote by $\bwt(i)$, $\bpsit(i)$, and $\bw^\star$ the block weight error vector, the block intermediate weight error vector, and the block optimum weight vector, all of size $N \times 1$ with blocks of size $L \times 1$, namely,
\begin{eqnarray}
	\bwt(i)	&\triangleq&\text{col}\{\bwt_1(i),\ldots,\bwt_N(i)\}		\\
	\bpsit(i)	&\triangleq&\text{col}\{\bpsit_1(i),\ldots,\bpsit_N(i)\}		\\
	\bw^\star	&\triangleq&\text{col}\{\bw_1^\star,\ldots,\bw_N^\star\}.
\end{eqnarray}
We also introduce the following $N\times N$ block matrices with individual entries of size $L\times L$:
\begin{eqnarray}
	\cb{\cp{M}}(i)	&\triangleq&\bM(i)\otimes \bI_L	\\
	\cb{\cp{A}}(i)	&\triangleq&\bA(i)\otimes \bI_L	\\
	\cb{\cp{P}}(i)	&\triangleq&\bP(i)\otimes \bI_L.
\end{eqnarray}
To perform the theoretical analysis, we introduce the following independence assumption.

{\assumption{(Independent regressors)
\label{Independent regressors} The regression vectors $\bx_k(i)$ arise from a stationary random process that is temporally stationary, temporally white, and independent over space with $\bR_{x,k} = E\{\bx_k(i)\,\bx^\top_k(i)\}>0$.}}

\medskip

\noindent A direct consequence is that $\bx_k(i)$ is independent of $\bwt_\ell(j)$ for all $\ell$ and $j \leq i$. Although not true in general, this assumption is commonly used to analyze adaptive constructions since it allows to simplify the derivations without constraining the conclusions. There are several results in the adaptation literature that show that performance results that are obtained under the above independence assumptions match well the actual performance of the algorithms when the step-sizes are sufficiently small (see, e.g.,~\cite[App. 24.A]{sayed2008adaptive} and the many references therein).

The estimation error in the first step of the asynchronous strategy \eqref{asynchronous_ATC} can be rewritten as:
\begin{equation}
	d_k(i)-\bx_k^\top(i)\bw_k(i)=\bx_k^\top(i)\bwt_k(i)+z_k(i).
\end{equation}
Subtracting $\bw^\star_k$ from both sides of the adaptation step in~\eqref{asynchronous_ATC} and using the above relation, we can express the update equation for $\bpsit(i+1)$ as:
\begin{equation}
\begin{split}
	\label{intermediate error}
	\bpsit(i+1)=&[\bI_{NL}-\cb{\cp{M}}(i)(\cbp{R}_{x}(i)+\eta\,\cbp{Q}(i))]\bwt(i)-\\
	&\cb{\cp{M}}(i)\bp_{xz}(i)+\eta\,\cb{\cp{M}}(i)\cbp{Q}(i)\bw^\star
\end{split}
\end{equation}
where
\begin{equation}
	\cbp{Q}(i)\triangleq\bI_{NL}-\cb{\cp{P}}(i),
\end{equation}
while $\cbp{R}_{x}(i)$ is an $N\times N$ block matrix with individual entries of size $L\times L$ given by
\begin{equation}
	\cbp{R}_{x}(i)\triangleq\text{diag}\big\{\bx_1(i)\bx^\top_1(i),\ldots,\bx_N(i)\bx^\top_N(i)\big\},
\end{equation}
and $\bp_{xz}(i)$ is the $N \times 1$ block column vector with blocks of size $L \times 1$ defined as
\begin{equation}
	\bp_{xz}(i)\triangleq\col\{\bx_1(i)z_1(i),\ldots,\bx_N(i)z_N(i)\}.
\end{equation}
Subtracting $\bw^\star_k$ from both sides of the combination step in~\eqref{asynchronous_ATC}, we get the block weight error vector:
\begin{equation}
	\label{eq:wt2psit}
	\bwt(i+1)=\cb{\cp{A}}^\top(i)\,\bpsit(i+1).
\end{equation}
Substituting \eqref{intermediate error} into \eqref{eq:wt2psit} we find that the error dynamics of the asynchronous multitask diffusion strategy \eqref{asynchronous_ATC} evolves according to the following recursion:
\begin{equation}
\begin{split}
	\label{error vector}
	\bwt(i+1)=&\cb{\cp{A}}^\top(i)\big[\bI_{NL}-\cb{\cp{M}}(i)(\cbp{R}_{x}(i)+\eta\,\cbp{Q}(i))\big]\bwt(i)-\\
	&\cb{\cp{A}}^\top(i)\cb{\cp{M}}(i)\bp_{xz}(i)+\eta\,\cb{\cp{A}}^\top(i)\cb{\cp{M}}(i)\cbp{Q}(i)\bw^\star.
\end{split}
\end{equation}
For compactness of notation, we introduce the symbols:
\begin{eqnarray}
				\cbp{B}(i)		&\triangleq&\cb{\cp{A}}^\top(i)\big[\bI_{NL}-\cb{\cp{M}}(i)(\cbp{R}_{x}(i)+\eta\,\cbp{Q}(i))\big]	\\
	\label{eq: g(i)}	\bg(i)			&\triangleq&\cb{\cp{A}}^\top(i)\cb{\cp{M}}(i)\,\bp_{xz}(i)							\\
	\label{eq: r(i)}    \br(i)			&\triangleq&\cb{\cp{A}}^\top(i)\cb{\cp{M}}(i)\bQ(i)\,\bw^\star,
\end{eqnarray}
so that \eqref{error vector} can be written as
\begin{equation}
	\label{error vector'}
	\bwt(i+1)=\cb{\cp{B}}(i)\,\bwt(i)-\bg(i)+\eta\br(i).
\end{equation}
Taking the expectation of both sides, using Assumption 1, and the independence of $\bA(i)$, $\bM(i)$, and $\bP(i)$, the network mean error vector ends up evolving according to the following dynamics:
\begin{equation}
	\label{eq:meanrec}
	\expec{\bwt(i+1)}={\cbp{B}}\,\expec{\bwt(i)}+\eta\br
\end{equation}
where
\begin{eqnarray}
	\label{eq:momentsBr}
	\cbp{B}	&\triangleq&\expec{\cbp{B}(i)}=\cbp{A}^\top\big[\bI_{NL}-\cbp{M}(\cbp{R}_x+\eta\cbp{Q})\big]	\\
	\br		&\triangleq&\expec{\br(i)}=\cbp{A}^\top\cbp{M}\cbp{Q}\,\bw^\star,
\end{eqnarray}
where $\cbp{A}$, $\cbp{M}$, $\cbp{R}_x$, and $\cbp{Q}$ denote the expectations of $\cbp{A}(i)$, $\cbp{M}(i)$, $\cbp{R}_x(i)$, and $\cbp{Q}(i)$, respectively, and are given by:
\begin{eqnarray}
	\cbp{A}	&\triangleq&\expec{\cbp{A}(i)}	=\overline{\bA}\otimes \bI_L	\\
	\cbp{M}	&\triangleq&\expec{\cbp{M}(i)}	=\overline{\bM}\otimes \bI_L	\\
	\cbp{P}	&\triangleq&\expec{\cbp{P}(i)}	=\overline{\bP}\otimes \bI_L	\\
	\cbp{R}_x &\triangleq&\mathbb{E}\{\cbp{R}_x(i)\}=\text{diag}\{\bR_{x,1},\ldots,\bR_{x,N}\}\\
	\label{eq: mean Q}\cbp{Q}	&\triangleq&\expec{\cbp{Q}(i)}	=\bI_{NL}-\expec{\cbp{P}(i)}	=\bI_{NL}-\cbp{P}.	
\end{eqnarray}
Note that $\expec{\bg(i)}=0$ since $z_k(i)$ is zero-mean and independent of any other signal.

{\theorem \textbf{(Stability in the mean)} Assume data model~\eqref{eq:datamodel} and Assumption 1 hold. Then, for any initial condition, the multitask diffusion LMS strategy \eqref{asynchronous_ATC} applied to asynchronous networks converges asymptotically in the mean if, and only if, the step-sizes in $\cbp{M}$ are chosen to satisfy
\begin{equation}
	\label{condition1}
	\rho\big(\cbp{A}^\top\big[\bI_{NL}-\cbp{M}(\cbp{R}_x+\eta\cbp{Q})\big]\big)<1,
\end{equation}  
where $\rho(\cdot)$ denotes the spectral radius of its matrix argument. In that case, the asymptotic mean bias is given by
\begin{equation}
	\label{mean ss}
	\lim_{i\rightarrow\infty}\expec{\bwt(i)}=\eta\,(\bI_{NL}-\cbp{B})^{-1}\br.
\end{equation}
Assume that the expected values for all step-sizes are uniform, namely, $\expec{\mu_k(i)}=\bar\mu$ for all $k$. A sufficient condition for \eqref{condition1} to hold is to ensure that
\begin{equation}
	\label{eq:condmean}
	0<\bar\mu<\frac{2}{\max_{1\leq k \leq N}\rho(\bR_{x,k})+2\eta}.
\end{equation}}

{\begin{proof}
Convergence in the mean requires the matrix $\cbp{B}$ in \eqref{eq:meanrec} to be stable. Since any induced matrix norm is lower bounded by its spectral radius, we can write in terms of the block maximum norm \cite{Sayed2013intr}:
\begin{align}
	&\rho\big(\cbp{A}^\top[\bI_{NL}-\cbp{M}(\cbp{R}_x+\eta\cbp{Q})]\big)\nonumber\\
		&\leq\|\cbp{A}^\top[\bI_{NL}-\cbp{M}(\cbp{R}_x+\eta\cbp{Q})]\|_{b,\infty}\nonumber\\
		&\leq\|\cbp{A}^\top\|_{b,\infty}\cdot\|\bI_{NL}-\cbp{M}(\cbp{R}_x+\eta\cbp{Q})\|_{b,\infty}.
\end{align}
We have $\|\cbp{A}^\top\|_{b,\infty}=1$ because $\cbp{A}$ is a block left-stochastic matrix. This yields:
\begin{align}
	&\rho\big(\cbp{A}^\top[\bI_{NL}-\cbp{M}(\cbp{R}_x+\eta\cbp{Q})]\big)\nonumber\\
		&\leq\|\bI_{NL}-\cbp{M}(\cbp{R}_x+\eta\cbp{Q})\|_{b,\infty}\nonumber\\
		&=\|\bI_{NL}-\cbp{M}(\cbp{R}_x+\eta(\bI_{NL}-\cbp{P}))\|_{b,\infty}\nonumber\\
		&\leq\|\bI_{NL}-\cbp{M}\cbp{R}_x-\eta\cbp{M}\|_{b,\infty}+\eta\| \cbp{M}\cbp{P}\|_{b,\infty}.\label{eq:rhomajo}
\end{align}
Consider the first term on the RHS of \eqref{eq:rhomajo}. Since the matrices $\cbp{M}$ and $\cbp{R}_x$ are block diagonal, it holds from the properties of the block maximum norm \cite{Sayed2013intr}:
\begin{align}
\label{eq: first term condition}
	&\|\bI_{NL}-\cbp{M}\cbp{R}_x-\eta\cbp{M}\|_{b,\infty}\nonumber\\	
		&=\max_{1\leq k\leq N} \rho\big((1-\eta\bar\mu_k) \bI_L-\bar\mu_k\bR_{x,k}\big) \notag\\
		&=\max_{1\leq k\leq N}\max_{1\leq \ell\leq L}|(1-\eta\bar\mu_k)-\bar\mu_k\lambda_\ell(\bR_{x,k})|
\end{align}
where $\bar\mu_k\triangleq\expec{\mu_k(i)}$, and $\lambda_\ell(\cdot)$ denotes the $\ell$-th eigenvalue of its matrix argument. Consider now the second term on the RHS of \eqref{eq:rhomajo}. Using the submultiplicative property of the block maximum norm,  and the fact that $\cbp{P}$ is a block right-stochastic matrix, we get
\begin{equation}
	\eta\|\cbp{M}\cbp{P}\|_{b,\infty}\leq\eta\|\cbp{M}\|_{b,\infty}.
\end{equation}
Because $\cbp{M}$ is a block diagonal matrix, we further have that
\begin{equation}
\label{eq: second term condition}
	\|\cbp{M}\|_{b,\infty}=\max_{1\leq k \leq N}\bar\mu_k.
\end{equation}
Combining \eqref{eq: first term condition} and \eqref{eq: second term condition} we conclude that the algorithm is stable in the mean if
\begin{equation}
	\label{eq:meanstab}
	\max_{1\leq k\leq N}\max_{1\leq \ell\leq L}|1-\eta\bar\mu_k-\bar\mu_k\lambda_{\ell}(\bR_{x,k})|
	+\eta\max_{1\leq k\leq N}\bar\mu_k < 1.
\end{equation}
In order to simplify this condition, assume that $\bar\mu_k=\bar{\mu}$ for all $k$. Condition~\eqref{eq:meanstab} then reduces to~\eqref{eq:condmean}. Note that the randomness in the topology does not affect the condition for stability in the mean of the algorithm.\end{proof}}

\subsection{Mean-square error behavior analysis}

To perform mean-square error analysis over asynchronous networks, compared to synchronous networks~\cite{chen2013multitask}, new operators with additional properties must be introduced. We shall use the block Kronecker product operator $\otimes_b$ instead of the Kronecker product $\otimes$, and the block vectorization operator $\text{bvec}(\cdot)$ instead of the vectorization operator $\text{vec}(\cdot)$. This is because, as explained in \cite{SayedPart2,sayed2014adaptation}, these block operators preserve the locality of the blocks in the original matrix arguments. Recall that if $\bX$ is an $N\times N$ block matrix with blocks of size $L\times L$, $\bvc(\bX)$ vectorizes each block of $\bX$ and stacks the vectors on top of each other. Before proceeding, we recall some properties of these block operators \cite{bvec,sayed2014adaptation}:\\

\noindent For any two $N\times 1$ block vectors $\{\bx,\by\}$ with blocks of size $L\times 1$, we have:
\begin{equation}
	\label{property 1}
	\text{bvec}(\bx\by^\top)=\by\otimes_b\bx.
\end{equation}

\noindent For any $N\times N$ block-matrices $\{\bA,\bB,\bC,\bD\}$ with blocks of size $L\times L$, we have:
\begin{equation}
	\label{property 2}
	(\bA+\bB)\otimes_b(\bC+\bD)=\bA\otimes_b \bC+\bA\otimes_b \bD+\bB\otimes_b \bC+\bB\otimes_b \bD
\end{equation}
\begin{equation}
	\label{property 3}
	(\bA\bC)\otimes_b(\bB\bD)=(\bA\otimes_b \bB)(\bC\otimes_b \bD)
\end{equation}
\begin{equation}
	\label{property 4}
	(\bA\otimes \bB)\otimes_b (\bC\otimes \bD)=(\bA\otimes \bC)\otimes(\bB\otimes \bD)
\end{equation}
\begin{equation}
	\label{property 5}
	\tr(\bA\bB)=[\,\bvc(\bB^\top)]^\top\text{bvec}(\bA)
\end{equation}
\begin{equation}
	\label{property 6}
	\text{bvec}(\bA\bB\bC)=(\bC^\top\otimes_b \bA)\,\text{bvec}(\bB)
	\end{equation}
\begin{equation}
	\label{property 7}
	(\bA\otimes_b \bB)^\top=(\bA^\top \otimes_b \bB^\top).
\end{equation}
We now use these properties to evaluate the expectation of some block Kronecker matrix products that will be useful in the sequel:
\begin{eqnarray}
	\label{expression M}
	\cbp{M}_{\text{\,I}}&\triangleq&\expec{\cbp{M}(i)\otimes_b\cbp{M}(i)}\nonumber\\
		&=&\mathbb{E}\{(\boldsymbol{M}(i)\otimes\bI_{L})\otimes_b(\boldsymbol{M}(i)\otimes\bI_{L})\} \notag\\
		&\overset{\eqref{property 5}}{=}&\mathbb{E}\{(\boldsymbol{M}(i)\otimes\boldsymbol{M}(i))\otimes(\bI_{L}\otimes\bI_{L})\} \notag\\
		&\overset{\eqref{eq:CM}}{=}&(\meanM\otimes\meanM+\bC_M)\otimes\bI_{L^2}.
\end{eqnarray}
In the same way, we get the following expectations: 
\begin{equation}
	\label{expression A}
	\cbp{A}_{\text{\,I}}\triangleq\expec{\cb{\cp{A}}(i)\otimes_b\cb{\cp{A}}(i)}
			=(\meanA\otimes\meanA+\bC_A)\otimes\bI_{L^2},
\end{equation}
\begin{equation}
	\label{expression P}
	\cbp{P}_{\text{\,I}}\triangleq\expec{\cbp{P}(i)\otimes_b\cbp{P}(i)}
			=(\meanP\otimes\meanP+\bC_P)\otimes\bI_{L^2}.
\end{equation}
Since $\cbp{Q}(i)=\bI_{NL}-\cbp{P}(i)$, we also obtain:
\begin{align}
	\label{expression Q}
	\cbp{Q}_{\text{\,I}}&\triangleq\expec{\cbp{Q}(i)\otimes_b\cbp{Q}(i)}\nonumber\\
		&=(\bI_{N^2}-\bI_{N}\otimes\overline{\bP}-\overline{\bP}\otimes\bI_N
		+\overline{\bP}\otimes\overline{\bP}+\textbf{\textit{C}}_P)\otimes\bI_{L^2}.
\end{align}
Before concluding these preliminary calculations, let us make some remarks on the stochasticity of matrices considered in the sequel. At each time instant $i$, the matrix $\bP(i)\otimes\bP(i)$ has nonnegative entries since $\bP(i)$ has nonnegative entries. It follows that  $\expec{\bP(i)\otimes\bP(i)}=\overline{\bP}\otimes\overline{\bP}+\bC_P$ has also nonnegative entries, and is right-stochastic since 
\begin{align}
	(\overline{\bP}\otimes\overline{\bP}+\bC_P)\cb{1}_{N^2}
	&=\expec{(\bP(i)\otimes\bP(i))(\cb{1}_N\otimes \cb{1}_N)}\nonumber\\
	&=\expec{(\bP(i)\,\cb{1}_N)\otimes(\bP(i)\,\cb{1}_N)}=\cb{1}_{N^2}
\end{align}
In the same token, the matrix $\overline{\bA}\otimes\overline{\bA}+\bC_A$ is left-stochastic.

To analyze the convergence in mean-square-error sense of the multitask diffusion LMS algorithm \eqref{asynchronous_ATC} over asynchronous networks, we consider  the variance of the weight error vector $\bwt(i)$ weighted by any positive semi-definite matrix $\bSig$, that is, $\expec{\|\bwt(i)\| ^2_{\bSig}}$, where $\|\bwt(i)\|^2_{\bSig}\triangleq\bwt^\top(i)\,\bSig\,\bwt(i)$. The freedom in selecting $\bSig$ will allow us to extract various types of information about the network and the nodes. By Assumption 1 and using \eqref{error vector'}, we get:
\begin{align}
	\label{eq:rec2order}
	\expec{\|\bwt(i+1)\| ^2_{\bSig}}=&\expec{\| \bwt(i)\| ^2_{\bSig' }}
	+\expec{\|\boldsymbol{g}(i)\|^2_{\bSig}}+\nonumber\\
	&\eta^2\expec{\|\br(i)\|^2_{\bSig}}+2\eta\expec{\br^\top(i)\bSig\cb{\cp{B}}(i)\bwt(i)}
\end{align}
where $\bSig'=\expec{\cbp{B}^\top(i)\bSig\cbp{B}(i)}$. Let $\bsig$ denotes the $(NL)^2\times 1$ vector representation of $\bSig$ that is obtained by the block vectorization operator, namely, $\bsig\triangleq\text{bvec}(\bSig)$. In the sequel, it will be more convenient to work with $\bsig$ than with $\bSig$ itself. Let $\bsig'\triangleq\text{bvec}(\bSig')$. Using property \eqref{property 6}, we can verify that
\begin{equation}
	\bsig'=\cbp{F}^\top\bsig
\end{equation}
where $\cbp{F}$ is the $(NL)^2\times(NL)^2$ matrix given by:
\begin{align}
\label{eq: expression of F}
	\cbp{F}	&\hspace{-1mm}\triangleq\expec{\cbp{B}(i)\otimes_b\cbp{B}(i)}\notag \\
			&\hspace{-2mm}\overset{\eqref{property 3}}{=}\expec{\cbp{A}^\top(i)\otimes_b\cbp{A}^\top(i)}\expec{[\bI_{NL}-\cbp{M}(i)(\cbp{R}_x(i)+\eta\cbp{Q}(i))]\nonumber\\
			&~\quad\quad\quad\quad\quad\quad\quad\quad\otimes_b\hspace{-1mm}[\bI_{NL}-\cbp{M}(i)(\cbp{R}_x(i)+\eta\cbp{Q}(i))]}\notag\\
			&\hspace{-4.5mm}\overset{\eqref{expression A},\eqref{property 2}}{=}\hspace{-1.5mm}\cbp{A}^\top_\text{\,I}[\bI_{(NL)^2}-\bI_{NL}\otimes_b\cbp{M}(\cbp{R}_x+\eta\cbp{Q})-\nonumber\\
			&\quad\quad\quad\cbp{M}(\cbp{R}_x+\eta\cbp{Q})\otimes_b\bI_{NL}+\notag\\
			&\quad\expec{\cbp{M}(i)(\cbp{R}_x(i)+\eta\cbp{Q}(i))\otimes_b\cbp{M}(i)(\cbp{R}_x(i)+\eta\cbp{Q}(i))}]
\end{align}
where using property \eqref{property 3} and the definition of $\cbp{M}_{\text{I}}$ in \eqref{expression M}, we have
\begin{align}
	\label{eq: expectation depending on square of step size}
	&\expec{\cbp{M}(i)(\cbp{R}_x(i)+\eta\cbp{Q}(i))\otimes_b\cbp{M}(i)(\cbp{R}_x(i)+\eta\cbp{Q}(i))}\nonumber\\
	&=\cbp{M}_{\text{\,I}}\,\expec{(\cbp{R}_x(i)+\eta\cbp{Q}(i))\otimes_b(\cbp{R}_x(i)+\eta\cbp{Q}(i))}.
\end{align}
The term on the RHS of equation~\eqref{eq: expectation depending on square of step size} is proportional to $\cbp{M}_{\text{\,I}}=\expec{\bM(i)\otimes \bM(i)}\otimes\bI_{L^2}$, where $\expec{\bM(i)\otimes \bM(i)}$ is an $N\times N$ block diagonal matrix whose $k$-th block is an $N\times N$ diagonal matrix with $\ell$-th entry given by $\expec{\mu_k(i)\mu_{\ell}(i)}$. It is sufficient for the exposition in this work to focus on the case of sufficiently small step-sizes where terms involving higher order moments of the step-sizes can be ignored. Such approximations are common when analyzing diffusion strategies in the mean-square-error sense (see \cite[Section 6.5]{Sayed2013intr}). Accordingly, the last term in \eqref{eq: expression of F} can be neglected and we continue our discussion by letting
\begin{align}
	\label{eq:approxF}
	\cbp{F}\approx\cbp{A}_{\text{\,I}}^\top[&\bI_{(NL)^2}-\bI_{NL}\otimes_b\cbp{M}(\cbp{R}_x
			+\eta\cbp{Q})-\nonumber\\&\cbp{M}(\cbp{R}_x+\eta\cbp{Q})\otimes_b\bI_{NL}].
\end{align}
Consider next the second term on the RHS of \eqref{eq:rec2order}. We can write:
\begin{equation}
	\expec{\|\boldsymbol{g}(i)\|^2_{\bSig}} = \tr\big\{\bSig\,\expec{\bg(i)\,\bg^\top(i)}\big\} \overset{\eqref{property 5}}{=} \bg_b^\top\bsig
\end{equation}
where $\bg_b=\bvc(\expec{\bg(i)\,\bg^\top(i)})$. Using expression \eqref{eq: g(i)} and the definitions of $\cbp{M}_{\text{\,I}}$ and $\cbp{A}_{\text{\,I}}$ in \eqref{expression M} and \eqref{expression A}, we have
\begin{align}
\label{eq: expression g_b}
	&\bg_b\hspace{-1mm}=\bvc(\expec{(\cb{\cp{A}}^\top(i)\cb{\cp{M}}(i)\,\bp_{xz}(i)\,\bp_{xz}^\top(i)\cb{\cp{M}}(i)\cb{\cp{A}}(i))}\notag\\
	&\hspace{1.75mm}\overset{\eqref{property 6}}{=}\expec{(\cbp{A}^\top(i)\otimes_b\cbp{A}^\top(i))\,\bvc\,(\cb{\cp{M}}(i)\,\bp_{xz}(i)\,\bp_{xz}^\top(i)\cb{\cp{M}}(i))}\notag \\
	&\hspace{-0.65mm}\overset{\eqref{property 6},\eqref{property 7}}{=}\cbp{A}_{\text{\,I}}^\top\,\expec{(\cbp{M}(i)\otimes_b\cbp{M}(i))\,\bvc\,(\bp_{xz}(i)\,\bp_{xz}^\top(i))}\notag \\
	&\hspace{2.5mm}=\cbp{A}_{\text{\,I}}^\top\cbp{M}_{\text{\,I}}\,\text{bvec}(\cbp{S}),
\end{align}
where $\cbp{S}\triangleq\expec{\bp_{xz}(i)\bp_{xz}^\top(i)}=\text{diag}\{\sigma^2_{z,k}\bR_{x,k}\}_{k=1}^N$. Let us examine now the third term on the RHS of \eqref{eq:rec2order}:
\begin{equation}
	\expec{\|\br(i)\|^2_{\bSig}}=\tr\big\{\bSig\,\expec{\br(i)\,\br^\top(i)}\big\}\overset{\eqref{property 5}}{=}\br_b^\top\bsig
\end{equation}
where $\br_b=\bvc(\expec{\br(i)\,\br^\top(i)})$. Using expression \eqref{eq: r(i)}, property \eqref{property 6}, and the definitions of $\cbp{M}_{\text{\,I}}\,$, $\cbp{A}_{\text{\,I}}$, and $\cbp{Q}_{\text{\,I}}$ in \eqref{expression M}, \eqref{expression A}, and \eqref{expression Q}, and proceeding as in \eqref{eq: expression g_b}, we obtain the following expression:
\begin{equation}
\br_b=\cbp{A}_{\text{\, I}}^\top\cbp{M}_{\text{\, I}}\cbp{Q}_{\text{\, I}}\,\bvc\,(\bw^\star(\bw^\star)^\top).
\end{equation}
Consider now the fourth term $\expec{\br^\top(i)\bSig\cb{\cp{B}}(i)\bwt(i)}$. We have:
\begin{eqnarray}
	\expec{\br^\top(i)\bSig\cb{\cp{B}}(i)\bwt(i)}
	&=&	\expec{\bvc(\br^\top(i)\bSig\cb{\cp{B}}(i)\bwt(i))}\notag\\
	&\overset{\eqref{property 6}}{=}&\expec{(\cb{\cp{B}}(i)\bwt(i))^\top\otimes_b\br^\top(i)}\,\bsig\notag \\
	&\overset{\eqref{property 7}}{=}&\expec{\cb{\cp{B}}(i)\bwt(i)\otimes_b\br(i)}^\top\bsig \notag\\
	&\overset{\eqref{property 3}}{=}&\expec{\bwt(i)\otimes_b {1}}^ \top\expec{\cb{\cp{B}}(i)\otimes_b\br(i)}^\top\bsig\notag\\
	&=&\expec{\bwt(i)}^\top\expec{\cb{\cp{B}}(i)\otimes_b\br(i)}^\top\bsig
\end{eqnarray}
with
\begin{align}
\label{eq: getting the fourth term 1}
	&\expec{\cbp{B}(i)\otimes_b\br(i)}\nonumber\\
	&=\mathbb{E} \{\cb{\cp{A}}^\top(i)\big[\bI_{NL}-\cb{\cp{M}}(i)(\cbp{R}_{x}(i)+\eta\,\bQ(i))\big]\otimes_b\nonumber\\
	&\quad\quad\quad\quad\quad\quad\quad\quad\quad\cb{\cp{A}}^\top(i)\cb{\cp{M}}(i)\bQ(i)\,\bw^\star\}\notag\\
	&\overset{\eqref{property 3}}{=}\cbp{A}_{\text{\,I}}^\top\,\mathbb{E}\{\big[\bI_{NL}-\cb{\cp{M}}(i)(\cbp{R}_{x}(i)+\eta\,\bQ(i))\big]\otimes_b\nonumber\\
	&\quad\quad\quad\quad\quad\quad\quad\quad\quad\cb{\cp{M}}(i)\bQ(i)\,\bw^\star\}\notag\\ 
	&\overset{\eqref{property 2}}{=}\cbp{A}_{\text{\,I}}^\top\big[\big(\bI_{NL}\otimes_b\cbp{M}\cbp{Q}\bw^\star)-\nonumber\\
	&\quad\quad\quad\expec{\cb{\cp{M}}(i)(\cbp{R}_{x}(i)+\eta\,\bQ(i))\otimes_b\cb{\cp{M}}(i)\bQ(i)\,\bw^\star}\big],
\end{align}
where 
\begin{align}
\label{eq: getting the fourth term 2}
	&\expec{\cb{\cp{M}}(i)(\cbp{R}_{x}(i)+\eta\,\bQ(i))\otimes_b\cb{\cp{M}}(i)\bQ(i)\,\bw^\star}\nonumber\\
	&\overset{\eqref{property 3}}{=}\cbp{M}_{\text{\,I}}\,\expec{(\cbp{R}_{x}(i)+\eta\,\bQ(i))\otimes_b\bQ(i)\,\bw^\star}\notag\\
	&\overset{\eqref{property 2}}{=}\cbp{M}_{\text{\,I}}\,((\cbp{R}_{x}\otimes_b\bQ\,\bw^\star)+\eta\,\expec{\bQ(i)\otimes_b\bQ(i)\,\bw^\star})\notag\\
	&\overset{\eqref{property 3}}{=}\cbp{M}_{\text{\,I}}\,((\cbp{R}_{x}\otimes_b\bQ\,\bw^\star)+\eta\,\cbp{Q}_{\text{\,I}}(\bI_{NL}\otimes_b\bw^\star)\,).
\end{align}
Finally, combining \eqref{eq: getting the fourth term 1} and \eqref{eq: getting the fourth term 2} and introducing the notation $\cbp{K}$, we get
\begin{align}
\label{eq: getting the fourth term 3}
	\cbp{K}&\triangleq\expec{\cbp{B}(i)\otimes_b\br(i)}\nonumber\\
	&=\cbp{A}_{\text{\,I}}^\top\big[(\bI_{NL}\otimes_b\cbp{M}\cbp{Q}\bw^\star)-\cbp{M}_{\text{\,I}}
	\big((\cbp{R}_x\otimes_b\cbp{Q}\bw^\star)+\nonumber\\
	&\quad\quad\quad\quad\quad\eta\cbp{Q}_{\text{\,I}}(\bI_{NL}\otimes_b\bw^\star)\big)\big].
\end{align}
Relation~\eqref{eq:rec2order} can be written in a more compact form as
\begin{equation}
	\label{eq:rec2order-approx}
	\expec{\|\bwt(i+1)\| ^2_{\bsig}}=\expec{\|\bwt(i)\| ^2_{\cbp{F}^\top\bsig}}+\by^\top(i)\bsig,
\end{equation}
where $\by(i)$ is the $(LN)^2\times 1$ vector given by:
\begin{equation}
	\by(i)\triangleq\bg_b+\eta^2\,\br_b+2\eta\,\cbp{K}\expec{\bwt(i)}.
\end{equation}
In the sequel, we shall use the notations $\|\!\cdot\!\|_{\bSig}$ and $\|\!\cdot\!\|_{\bsig}$ interchangeably.

{\theorem \textbf{(Mean-square stability)} Assume data model \eqref{eq:datamodel} and Assumption 1 hold. Assume further that the upper bounds on the step-sizes, $\{\mu_{\max,k}\}$, are sufficiently small such that approximation~\eqref{eq:approxF} is justified by ignoring higher-order powers of the step-sizes, and \eqref{eq:rec2order-approx} can be used as a reasonable representation for the dynamics of the weighted mean-square error. Then, the asynchronous diffusion multitask algorithm~\eqref{asynchronous_ATC} is mean-square stable if the matrix $\cbp{F}$ defined by \eqref{eq:approxF} is stable.}

{\begin{proof}
Provided that $\cbp{F}$ is stable, recursion~\eqref{eq:rec2order-approx} is stable if $\by^\top(i)\bsig$ is bounded. Since $\eta,\bg_b,\br_b,\cbp{K}$, and $\bsig$ are finite and constant terms, the boundedness of $\by^\top(i)\bsig$ depends on $\expec{\bwt(i)}$ being bounded. We know from~\eqref{eq:meanrec} that $\expec{\bwt(i)}$ is uniformly bounded because \eqref{eq:meanrec} is a Bounded Input Bounded Output (BIBO) stable recursion with a bounded driving term $\eta\cbp{A}^\top\cbp{M}\cbp{Q}\,\bw^\star$.
It follows that $\by^\top(i)\bsig$ is uniformly bounded. As a result, $\mathbb{E}\{\|\bwt(i+1)\| ^2_{\bsig}\}$ converges to a bounded value as $i\rightarrow\infty$, and the algorithm is mean-square stable. 

The stability of $\cbp{F}$ is studied in Appendix \ref{stability of F}. It is worth noting that, due to the Kronecker covariance matrix $\bC_A$, the matrix $\cbp{F}$ cannot be approximated by $\cbp{B}\otimes\cbp{B}$ as in the synchronous case~\cite{Sayed2013intr,chen2013multitask}. Moreover, deriving a condition that ensures the stability of $\cbp{F}$ in a multitask setting is more challenging than in the single-task setting~\cite{SayedPart2} due to the presence of the non-block diagonal matrix $\cbp{Q}$ in the second term on the RHS of~\eqref{eq:approxF}.  
\end{proof}}

{\theorem\label{th:transient} \textbf{\textit{(Transient network performance)}} Consider sufficiently small step-sizes that ensure mean and mean-square stability. The variance curve defined by  $\zeta(i)=\expec{\|\bwt(i+1)\|^2_{\bsig}}$ evolves according to the following recursion for $i \geq 0$:
\begin{align}
	\label{Transient MSD1}
	&\zeta(i+1)\nonumber\\
	&=\zeta(i)+\|\bwt(0)\| ^2_{(\cbp{F}^\top-\bI_{(NL)^2})(\cb{\cp{F}}^\top)^{i}\bsig}+
	(\by^\top(i)+\bGam(i))\bsig
\end{align}
where $\boldsymbol{\Gamma}(i+1)$ is updated as follows:
\begin{equation}
	\label{Transient MSD2}
	\bGam(i+1)=\bGam(i)\,\cbp{F}^\top+\by^\top(i)(\cbp{F}^\top-\bI_{(NL)^2}),
\end{equation}
with the initial conditions $\zeta(0)=\|\bwt(0)\|^2_{\bsig}$ and $\bGam(0)=\cb{0}_{(NL)^2}^\top$.} The network mean-square deviation (MSD) is obtained by setting $\bsig=\bvc(\bSig)$ with $\bSig=\frac{1}{N}\bI_{NL}$.

{\begin{proof} The argument is similar to the proof of Theorem $3$ in \cite{chen2013multitask}. 
\end{proof}}

{\theorem\label{th:steady} \textbf{(\textit{Steady-state network performance})} Assume sufficiently small step-sizes to ensure mean and mean-square convergence. Then, the steady-state performance for multitask diffusion LMS \eqref{asynchronous_ATC} applied to asynchronous network is given by:
\begin{equation}
	\label{Steady State MSD}
	\zeta^\star=\big(\bg_b+\eta^2\,\br_b+2\eta\,\cbp{K}\,\expec{\bwt(\infty)}\big)^\top(\bI_{(NL)^2}-\cbp{F}^\top)^{-1}\bsig.
\end{equation}
where $\mathbb{E}\{\bwt(\infty)\}$ is given by (\ref{mean ss}). The network mean-square deviation (MSD) is obtained by setting $\bsig=\bvc(\bSig)$ with $\bSig=\frac{1}{N}\bI_{NL}$.}

{\begin{proof}
The steady-state network performance with metric $\bsig$ is defined as:
\begin{equation}
	\label{expression 1}
	\zeta^\star=\lim_{i\rightarrow\infty}\mathbb{E}\{\|\bwt(i)\|^2_{\bsig}\}.
\end{equation}
From the recursive expression \eqref{eq:rec2order-approx}, we obtain as $i\rightarrow \infty$:
\begin{align}
	\label{expression 2}
	&\lim_{i\rightarrow \infty}\expec{\|\bwt(i)\|^2_{(\bI_{(NL)^2}-\cb{\cp{F}}^\top)\bsig}}\nonumber\\
	&= (\bg_b+\eta^2\,\br_b+2\eta\,\cbp{K}\expec{\bwt(\infty)})^\top\bsig.
\end{align}
To obtain \eqref{expression 1}, we replace $\bsig$ in \eqref{expression 2} by $(\bI_{(NL)^2}-\cbp{F}^\top)^{-1}\bsig$.
\end{proof}}

Before moving on to the presentation of experimental results, note that the performance of the \textit{synchronous} multitask algorithm over the \textit{mean-graph} topology can be obtained by setting $\bC_A$, $\bC_M$, and $\bC_P$ to zero in \eqref{expression M}--\eqref{expression P}.

\section{Simulation results}
\label{simulation}
\subsection{Illustrative example}
\label{ss:simulation1}
\begin{figure}[!h]
	\centering
	\includegraphics[width=0.24\textwidth,height=0.17\textheight]{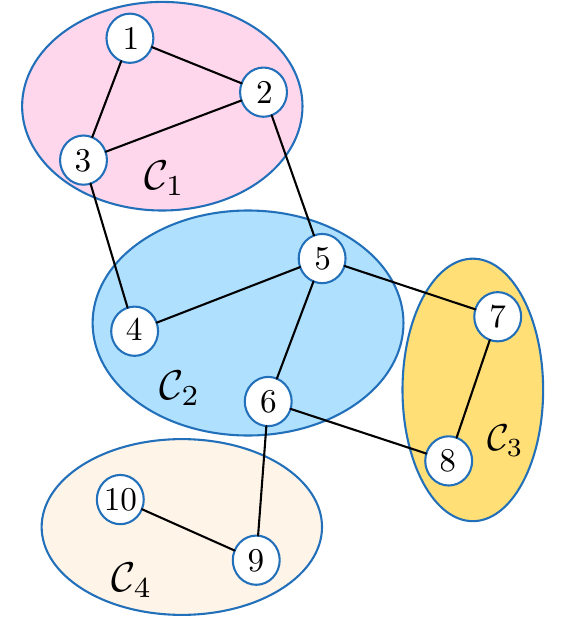}
	\includegraphics[width=0.242\textwidth,height=0.17\textheight]{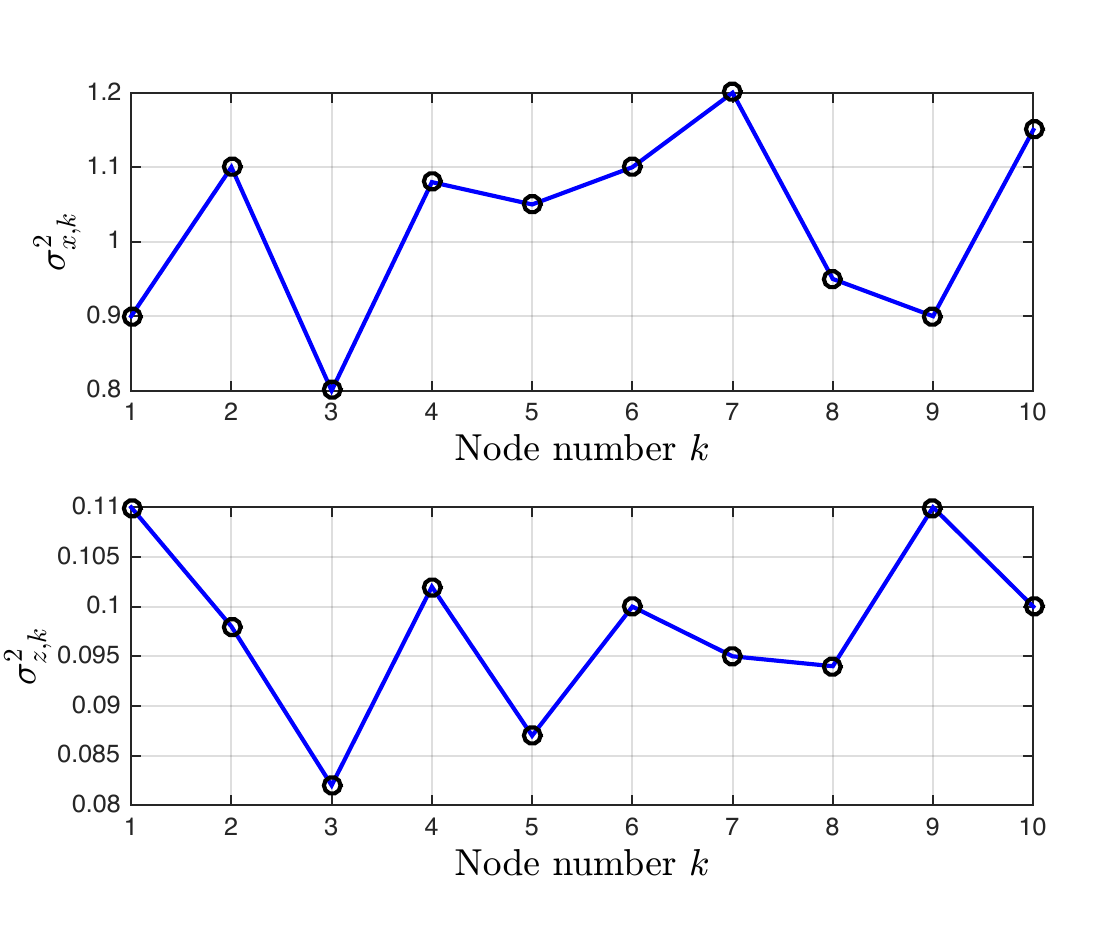}
	\caption{Experimental setup. Left: Network topology. Right: Regression and noise variances.}
	\label{fig:topology&data variances}
\end{figure}

We adopt the same clustered multitask network as \cite{chen2013multitask} in our simulations. As shown in Figure~\ref{fig:topology&data variances}, the network consists of $10$ nodes divided into $4$ clusters: $\C_1=\{1,2,3\}$, $\C_2=\{4,5,6\}$, $\C_3=\{7,8\}$, $\C_4=\{9,10\}$. The unknown parameter vector $\bw^\star_{\C_i}$ of each cluster is of size $2\times 1$, and has the following form: $\bw^\star_{\C_i}=\bw_0+\delta\bw_{\C_i}$ with $\bw_0=[0.5,-0.4]^\top$, $\delta\bw_{\C_1}=[0.0287,-0.005]^\top$, $\delta\bw_{\C_2}=[0.0234,0.005]^\top$, $\delta\bw_{\C_3}=[-0.0335,0.0029]^\top$, and $\delta\bw_{\C_4}=[0.0224,0.00347]^\top$. The input and output data at each node $k$ are related via the linear regression model: $d_k(i)=\bx^\top_{k}(i)\bw_k^\star+z_k(i)$ where $\bw_k^\star=\bw_{\C(k)}^\star$. The regressors are zero-mean $2\times 1$ random vectors governed by a Gaussian distribution with covariance matrices $\bR_{x,k}=\sigma^2_{x,k}\bI_{L}$. The variances $\sigma^2_{x,k}$ are shown in Figure~\ref{fig:topology&data variances}. The background noises $z_k(i)$ are independent and identically distributed zero-mean Gaussian random variables, independent of any other signals. The corresponding variances are given in Figure~\ref{fig:topology&data variances}.

\begin{figure*}[t]
\centering
 \includegraphics[trim = 15mm 80mm 20mm 80mm, clip, scale=0.4]{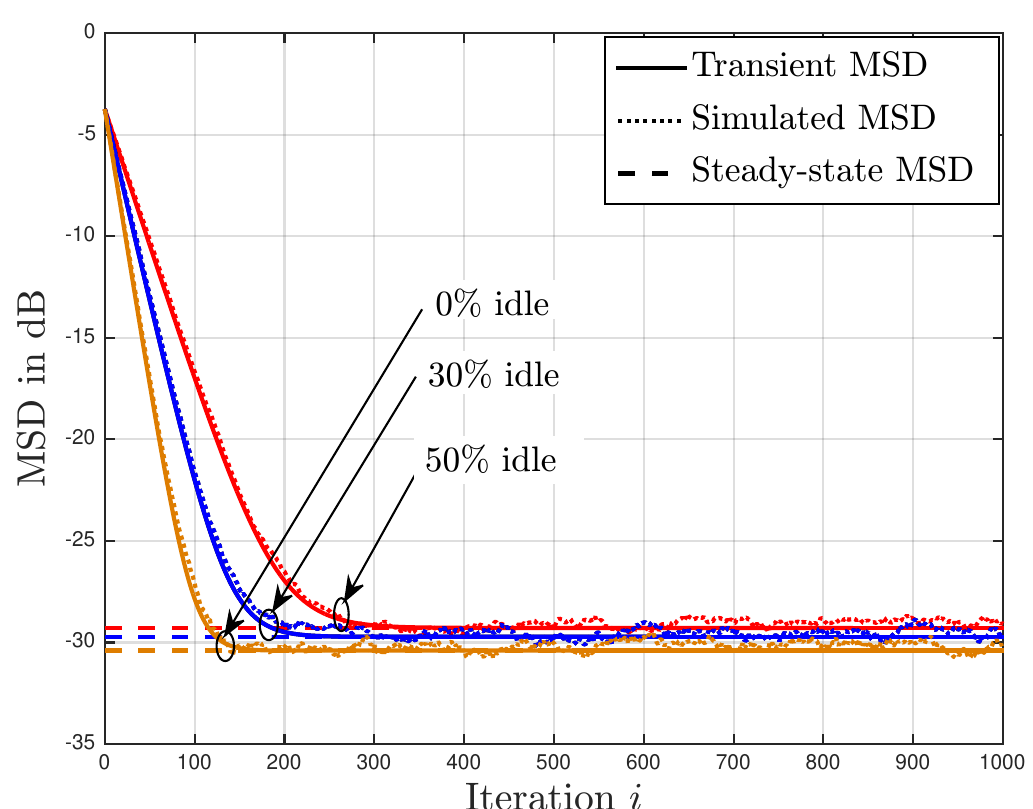}
  \includegraphics[trim =15mm 80mm 20mm 80mm, clip, scale=0.4]{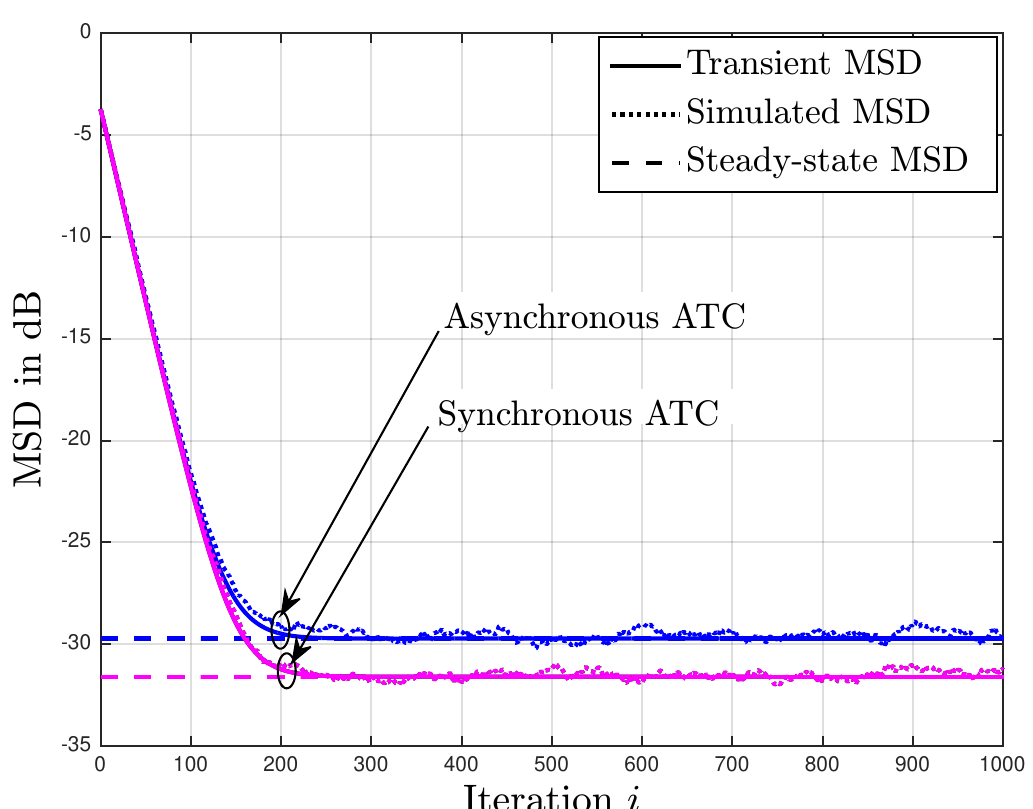}
\caption{Left: Comparison of asynchronous network MSD under $50\%$ idle, $30\%$ idle, and $0\%$ idle. Right: Network MSD comparison of asynchronous network under $30\%$ idle and the corresponding synchronous network.}
\label{fig:simulations1}
\end{figure*}

We considered the Bernoulli asynchronous model described in Appendix \ref{bernoulli model}. We set the coefficient $a_{\ell k}$ in~\eqref{eq: PD a} such that $a_{\ell k}=|\N_{k}\cap \C(k)|^{-1}$ for all $\ell\in(\N_{k}\cap\C(k))$, where $| \N_{k}\cap\C(k)|$ denotes the cardinality of the set $\N_{k}\cap \C(k)$. Then we set the regularization factors $\rho_{k\ell}$ in \eqref{eq: PD rho} as follows. If $\N_k\setminus\C(k)\neq\varnothing$, $\rho_{k\ell}$ was set to $\rho_{k\ell}=|\N_{k}\setminus \C(k)|^{-1}$ for all $\ell\in \N_k\setminus \C(k)$, and to $\rho_{k\ell}=0$ for any other $\ell$. If $\N_k\setminus \C(k)=\varnothing$, these factors were set to $\rho_{kk}=1$ and to $\rho_{k\ell}=0$ for all $\ell\neq k$. This usually leads to asymmetrical regularization factors. The parameters of the Bernoulli distribution governing the step-sizes $\mu_k(i)$ were the same over the network, that is, we set $\mu_k$ in \eqref{eq: PD mu} to $0.03$ for all $k$. The regularization strength $\eta$ was set to $1$. The MSD learning curves were averaged over $100$ Monte-Carlo runs. The transient MSD curves were obtained with Theorem~\ref{th:transient}, and the steady-state MSD was estimated with Theorem~\ref{th:steady}. In  Figure~\ref{fig:simulations1} (left), we report the network MSD learning curves for $3$ different cases:

\noindent Case 1: $50\%$ idle: $q_k=p_{\ell k}=r_{k\ell}=0.5$;

\noindent Case 2: $30\%$ idle: $q_{k}=p_{\ell k}=r_{k\ell}=0.7$;

\noindent Case 3: no idle nodes: $q_{k}=p_{\ell k}=r_{k\ell}=1$. 

\noindent We observe that the simulation results match well the theoretical results. Furthermore, the performance of the network is influenced by the probability of occurrence of random events. In Figure~\ref{fig:simulations1} (right), the asynchronous algorithm in Case 2 is compared with its synchronous version obtained from \eqref{synchronous_ATC} by setting $\mu_k$, $a_{\ell k}$, and $\rho_{k \ell}$ to the expected values $\bar{\mu}_k=\expec{\mu_k(i)}$, $\bar{a}_{\ell k}=\expec{a_{\ell k}(i)}$, and $\bar{\rho}_{k\ell}=\expec{\rho_{k\ell}(i)}$, respectively. Although both algorithms show the same convergence rate, the asynchronous algorithm suffers from degradation in its MSD performance caused by the additional randomness throughout the adaptation process.


\subsection{Multitask learning benefit}
\label{ss: multitask benefit}

\begin{figure}[!h]
	\centering
	\includegraphics[trim = 0mm 70mm 0mm 25mm,clip,scale=0.25]{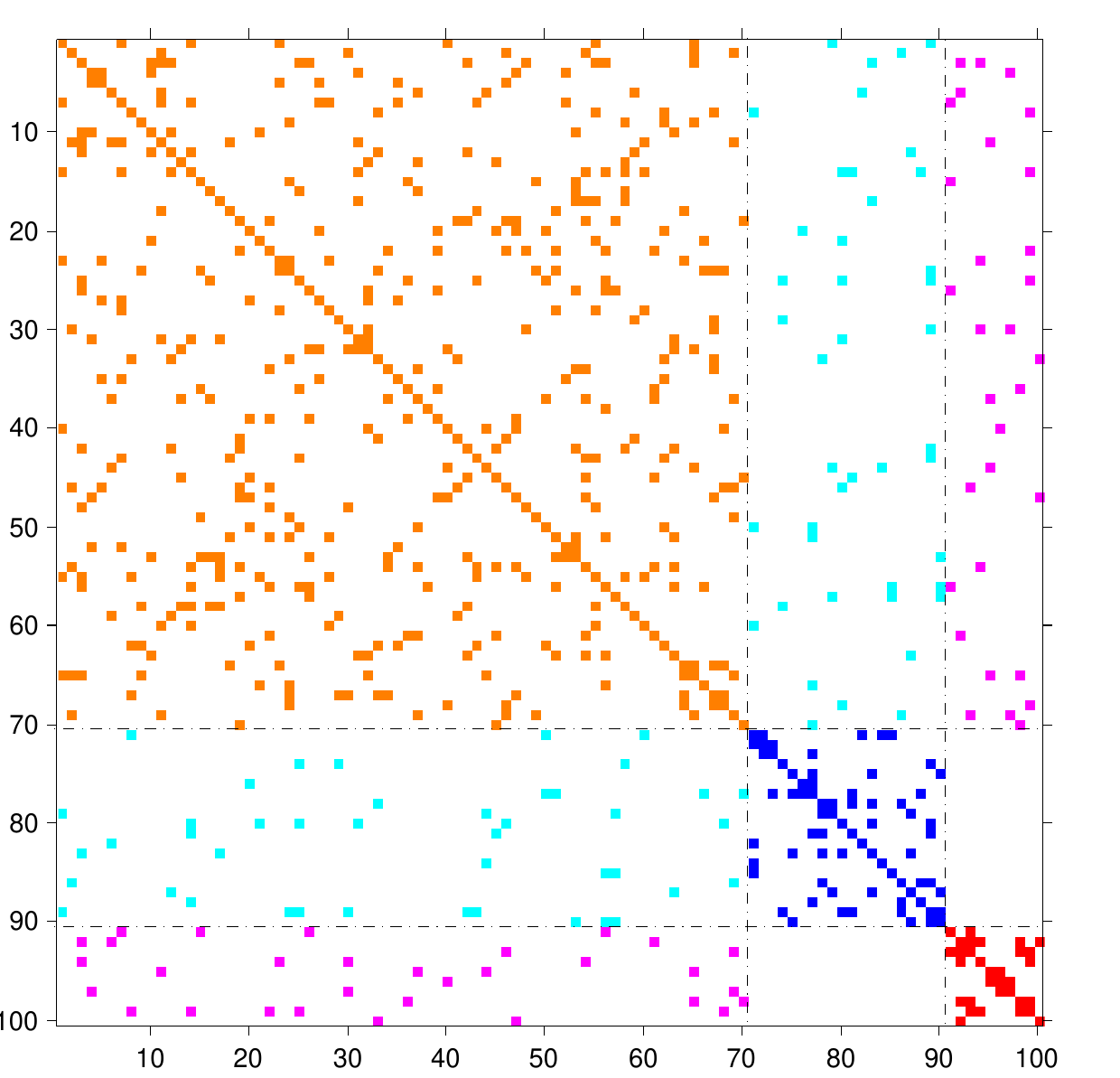}
	\caption{Connectivity matrix of the network. The orange, blue, and red elements correspond to links within $\C_1$, $\C_2$, and $\C_3$, respectively. The cyan elements correspond to links between $\C_1$ and $\C_2$ and the magenta elements correspond to links between $\C_1$ and $\C_3$. No links between $\C_2$ and $\C_3$.}
	\label{fig:topology2&datavariances}
\end{figure}

In this section we provide an example to show the benefit of multitask learning. We consider a network consisting of $N=100$ nodes grouped into $Q=3$ clusters such that $\C_1=\{1,\ldots,70\}$, $\C_2=\{71,\ldots,90\}$, and $\C_3=\{91,\ldots,100\}$. The physical connections are defined by the connectivity matrix represented in Figure \ref{fig:topology2&datavariances}. The inputs $\bx_k(i)$ were zero-mean $21\times 1$ random vectors governed by a Gaussian distribution with covariance matrix $\bR_{x,k}=\sigma^2_{x,k}\bI_{21}$, where $\sigma^2_{x,k}$ were randomly chosen in the interval $[1,1.4]$. The noises $z_k(i)$ were i.i.d. zero-mean Gaussian random variables, independent of any other signal with variances $\sigma^2_{z,k}$ randomly chosen in the interval $[0.1,0.15]$. The $21\times 1$ unknown parameter vectors were chosen as: $\bw^\star_{\C_1}=\bw_0=[\cb{1}_{1\times 3},\cb{0}_{1\times 3},2\cdot\cb{1}_{1\times 3},\cb{0}_{1\times 3},-\cb{1}_{1\times 3},\cb{0}_{1\times 3},-2\cdot\cb{1}_{1\times 3}]^\top$,  $\bw^\star_{\C_2}=\bw_0+\delta\bw$, $\bw^\star_{\C_3}=\bw_0-\delta\bw$ where $\delta\bw$ was randomly generated such that $\|\delta\bw\|_{\infty}=\max_i|[\delta\bw]_i|=0.03$. 

We considered the Bernoulli asynchronous model. The coefficients $\{a_{\ell k}\}$ and $\{\rho_{k\ell}\}$ in \eqref{eq: PD a} and \eqref{eq: PD rho}, respectively, were generated in the same manner as in \ref{ss:simulation1}. Parameters $\mu_k$ and $q_k$ in \eqref{eq: PD mu} were set to $\mu_k=1/30$, $q_k=0.8$ for nodes in the first cluster, $\mu_k=2/45$, $q_k=0.6$ for nodes in the second cluster, and $\mu_k=1/15$, $q_k=0.4$ for nodes in the third cluster. The probabilities $\{p_{\ell k}\}$ in \eqref{eq: PD a} were $p_{\ell k}=0.8$ for links in the first cluster, $p_{\ell k}=0.6$ for links in the second cluster, and $p_{\ell k}=0.4$ for links in the third cluster. The probability that a link connecting two nodes belonging to neighboring clusters drops was $1-r_{k\ell}=0.25$. The simulated curves were obtained by averaging over $150$ Monte-Carlo runs. 

\begin{figure*}[t]
\centering
               \includegraphics[trim = 18mm 80mm 13mm 80mm,scale=0.4]{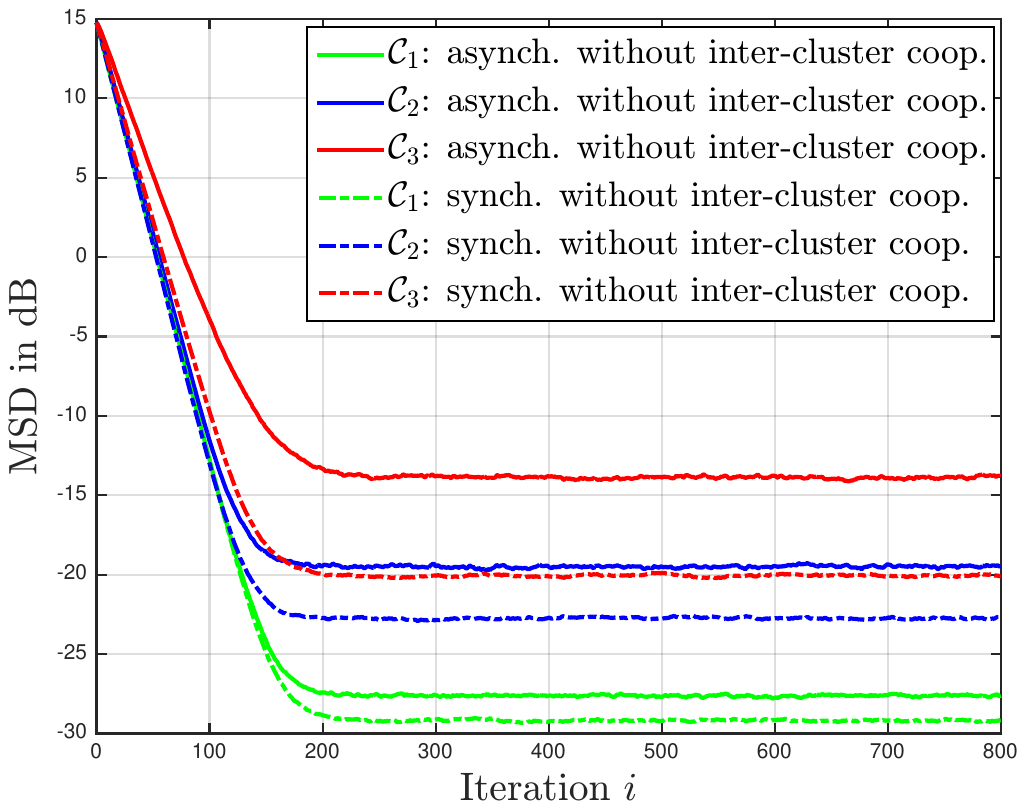}
               \includegraphics[trim = 18mm 80mm 15mm 80mm,scale=0.4]{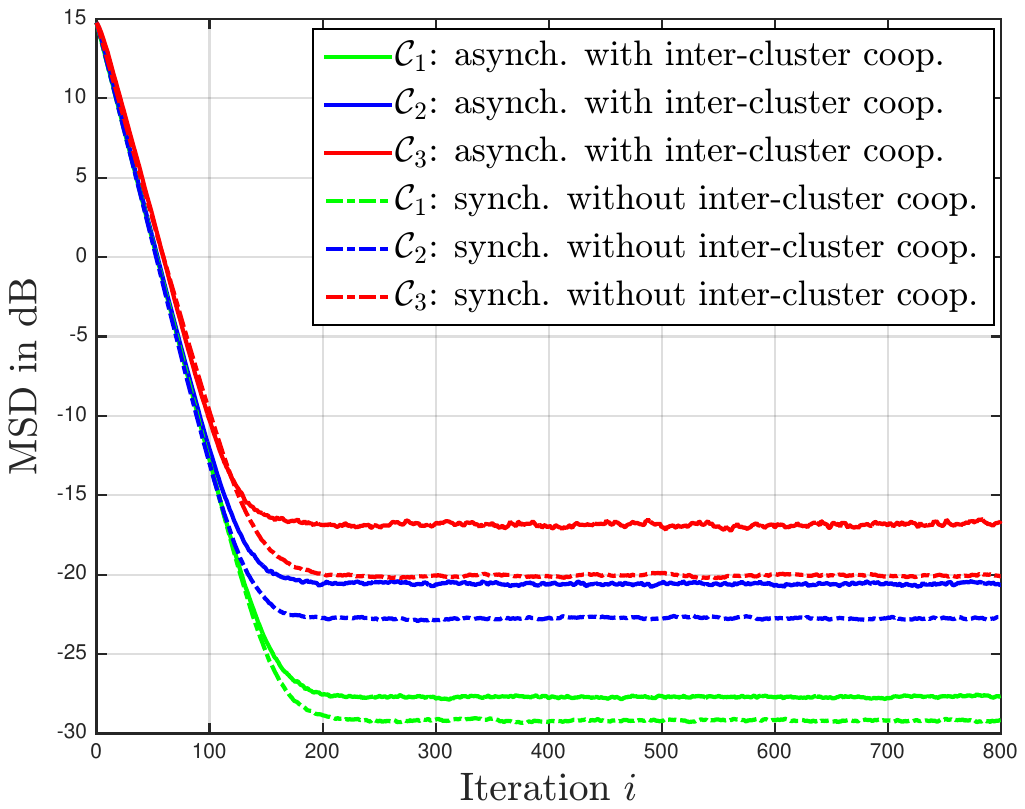}
      		\caption{Cluster learning curves. Left: Comparison of the asynchronous multitask diffusion LMS
		\eqref{asynchronous_ATC} without 
		inter-cluster cooperation $(\eta=0)$ and its synchronous counterpart. Right: Comparison of the asynchronous
		multitask diffusion LMS \eqref{asynchronous_ATC} with inter-cluster cooperation $(\eta \neq 0)$ and the
		multitask diffusion LMS \eqref{asynchronous_ATC} without
		inter-cluster cooperation $(\eta=0)$.}
			\label{fig:simulation2learningcurves}
\end{figure*}

In Figure \ref{fig:simulation2learningcurves} (left), we compare two algorithms: the asynchronous diffusion strategy without regularization (obtained from~\eqref{asynchronous_ATC} by setting $\eta=0$) and its synchronous counterpart (obtained from~\eqref{asynchronous_ATC} by setting $\eta=0$ and replacing $\mu_k(i),a_{\ell k}(i)$ by $\bar\mu_k,\bar{a}_{\ell k}$). As shown in this figure, the performance is highly deteriorated in the third cluster and slightly deteriorated in the first cluster because $\C_3$ is more susceptible to random events. In Figure \ref{fig:simulation2learningcurves} (right), we compare two algorithms: the asynchronous diffusion strategy with regularization (obtained from \eqref{asynchronous_ATC} by setting $\eta=2$) and the same synchronous algorithm as in the left plot. As shown in this figure, the cooperation between clusters improves the performance of each cluster so that gaps appearing in the left plot are reduced. In other words, $\C_2$ and $\C_3$ benefit from the high performance levels achieved by $\C_1$. This can be justified by two arguments: a large number of nodes is employed to collectively estimate $\bw^\star_{\C_1}$ and the probabilities associated with random events in $\C_1$ are small. As a conclusion, when tasks between neighboring clusters are similar, cooperation among clusters improves the learning especially for clusters where asynchronous events occur frequently.

\subsection{Circular arcs localization}
\label{ss: target localization}

In this section, we consider the problem of adaptive surface localization over asynchronous networks. When dealing with a smooth target surface, we can expect that promoting the smoothness of the graph signal  will improve the performance of the network~\cite{chen2013multitask}. In the following, we consider an arc localization application where the radius of the arc is changing over time, and we illustrate the influence of the random events on the learning behavior and tracking ability of the network. 

Let us denote by $\mathcal{L}=[\theta_0,\theta_1]$ an arc of circle with radius $R$ and subtending an angle $\theta=\theta_1-\theta_0$ with the circle center $\bw_o$. Let us decompose $\mathcal{L}$ into $Q$ sub-arcs $\mathcal{L}_q$ with radius $R$ and subtending an angle $\delta\ll\theta$ with $\bw_o$. In order to estimate the location of $\mathcal{L}$, and for sufficiently small $\delta$, it is sufficient to estimate the location of each of these $Q$ sub-arcs by solving a point target localization problem. This can be done by employing a network of $N$ nodes, composed of $Q$ clusters, where nodes of each cluster $\C_q$ are interested in locating $\mathcal{L}_q$ by estimating a parameter vector $\bw^\star_{\C_q}$. Let us consider node $k$ belonging to cluster $\C_q$. At each time instant $i$, node $k$ gets noisy measurements $\{d_k(i),\bu_k(i)\}$ that are related via the linear data model~\cite{Sayed2013intr}:
\begin{equation}
	d_k(i)=\bu_k^\top (i)\bw^\star_{\C_q}+v_k(i),
\end{equation}
where $v_k(i)$ is a zero-mean temporally and spatially independent Gaussian noise with variance $\sigma^2_{v,k}$, $\bu_k(i)$ is a noisy measurement of the  unit-norm direction vector of $\bu_k$ pointing from agent $k$ to the target $\bw^\star_{\C_q}$ given by:
\begin{equation}
	\bu_k(i)=\bu_k+\alpha_k(i)\bu_k^\perp+\beta_k(i)\bu_k,
\end{equation}
with $\bu_k$ given by $\bu_k=(\bw^\star_{\C_q}-\cb{n}_k)/\|\bw^\star_{\C_q}-\cb{n}_k\|$ where $\cb{n}_k$ is the location vector of node $k$, $\bu_k^\perp$ denoting a unit norm vector that lies in the same space as $\bu_k$ and whose direction is perpendicular to $\bu_k$. The variables $\alpha_k(i)$ and $\beta_k(i)$ are zero-mean independent Gaussian random variables of variances $\sigma^2_{\alpha,k}$ and $\sigma^2_{\beta,k}$, respectively. The amount of perturbation along the parallel direction is assumed to be small compared to the amount of perturbation along the perpendicular direction, that is,  $\sigma^2_{\beta,k}\ll\sigma^2_{\alpha,k}$.

\begin{figure}[!h]
	\centering
	\includegraphics[trim = 10mm 70mm 20mm 70mm,scale=0.35]{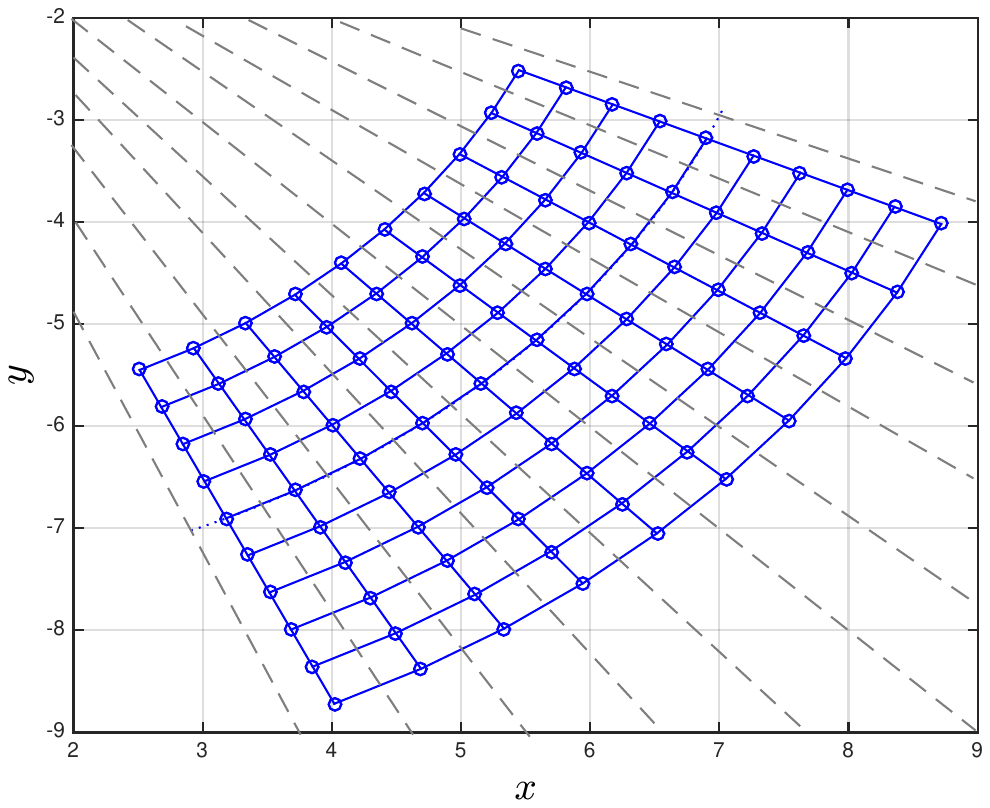}
	\caption{Network topology consisting of 10 clusters: circles for nodes, solid lines for links, and dashed lines for cluster boundaries.}
	\label{fig:topology3}
\end{figure}

To show the effects of randomness at the level of nodes and links, we considered a network of $100$ nodes grouped into $Q=10$ clusters, located over arcs of radiuses uniformly distributed between $3R_0$ and $5R_0$ given $R_0$. Angular parameters $\theta_0$ and $\theta_1$ were set to $13\pi/8$ and $15\pi/8$, respectively. The network topology is shown in Figure~\ref{fig:topology3}. The noise variances were set to $\sigma^2_{v,k}=0.2$, $\sigma^2_{\alpha,k}=0.05$, and $\sigma^2_{\beta,k}=0.005$, for all $k$. We considered a Bernoulli asynchronous model.  The coefficients $a_{\ell k}$ in~\eqref{eq: PD a} were set to $|\N_k\cap\C(k)|^{-1}$ for intra-cluster links, and to zero for inter-cluster links. The regularization factors $\rho_{k\ell}$ in \eqref{eq: PD rho} were set to $|\N_k\setminus\C(k)|^{-1}$.  The probabilities of success $q_k$, $p_{\ell k}$, and $r_{k \ell}$ were identically set to $0.5$.

\begin{figure*}[!h]
	\centering
	\includegraphics[scale=0.2]{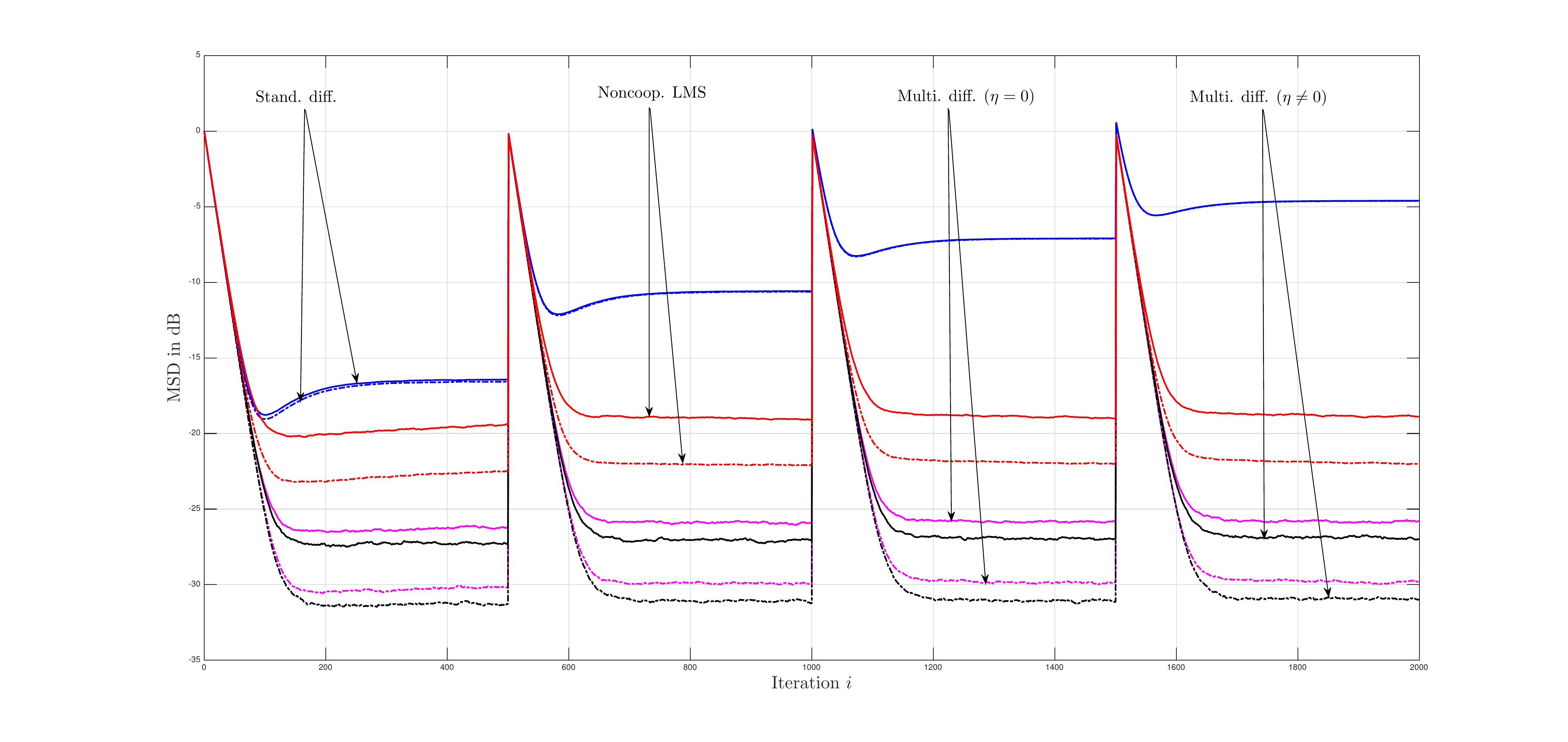}
	\caption{Network topology consisting of 10 clusters. Network MSD learning curves in a non-stationary environment: comparison of the multitask diffusion LMS with (namely, $\eta>0$) and without (namely, $\eta=0$) inter-cluster cooperation, the standard diffusion LMS~\cite{Lopes2008diff} and the non-cooperative LMS. The dotted lines are for synchronous networks and the solid lines are for asynchronous networks.}
	\label{fig:MSD comparison}
\end{figure*}

\begin{figure}[!h]
	\centering
	\includegraphics[trim = 10mm 70mm 20mm 80mm,scale=0.35]{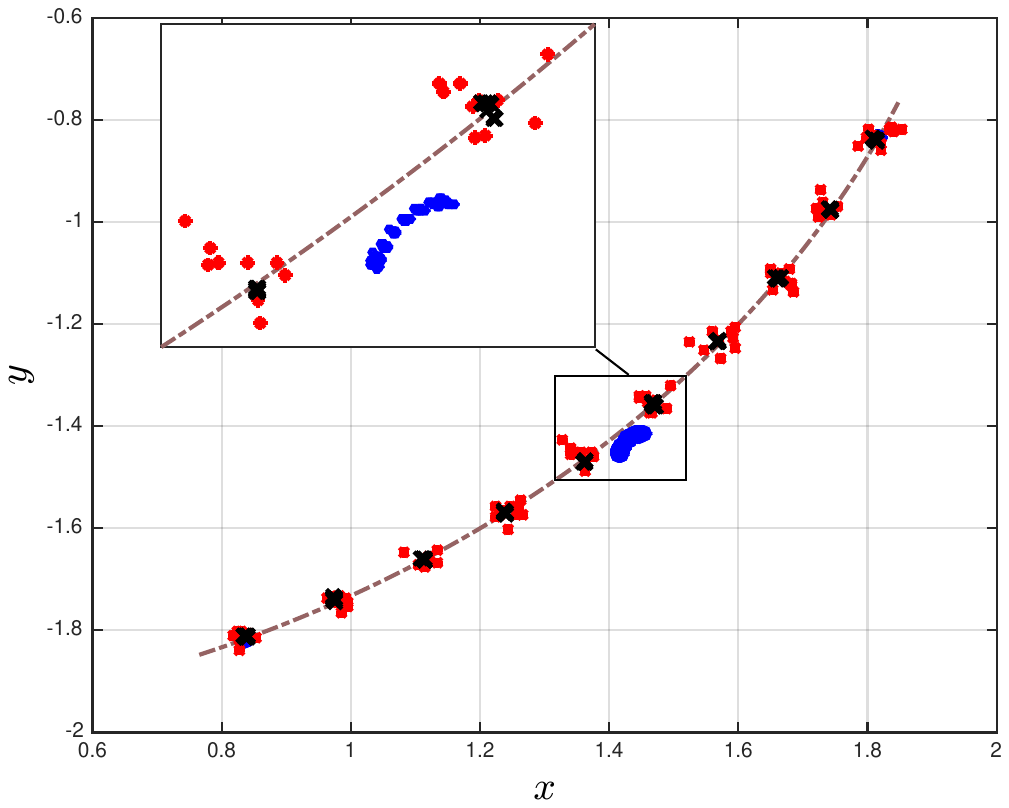}
	\caption{Target estimation results ($R=R_0=2$) over asynchronous network: black cross sign for multitask diffusion~\eqref{asynchronous_ATC}, red asterisk sign for non-cooperative, and blue circle sign for standard diffusion~\cite{Lopes2008diff}.}
	\label{fig:estimation comparison}
\end{figure}

The MSD learning curves were averaged over $200$ Monte-Carlo runs. We ran the synchronous and asynchronous multitask algorithms in two different situations. For the first one, we set the regularization strength $\eta$ to zero, that is, we did not allow any cooperation between neighboring clusters. In the second one, we set the regularization strength $\eta$ to $0.5$. For comparison purposes, we also ran the noncooperative LMS, which was obtained by setting $\bA(i)=\bP(i)=\bI_N$ for all $i$, and the standard diffusion LMS~\cite{Lopes2008diff}. In both cases, synchronous and asynchronous algorithms were also considered. Each synchronous algorithm was derived from its asynchronous counterpart by making $\mu_{k}(i)$, $a_{\ell k}(i)$, and $\rho_{k \ell}(i)$ deterministic quantities equal to $\bar{\mu}_{k}$, $\bar{a}_{\ell k}$ and $\bar{\rho}_{k \ell}$, respectively. In order to illustrate the tracking ability of the algorithms, we modified the radius $R$ of $\mathcal{L}$ every $500$ iterations such that: for $i\in[0,500]$, $R=0.5R_0$, for $i\in]500,1000]$, $R=R_0$, for $i\in]1000,1500]$, $R=1.5R_0$, and for $i\in]1500,2000]$, $R=2R_0$. Note that varying $R$ has an effect on the level of similarity between neighboring tasks when characterized by $\|\bw_{\C_i}^\star-\bw_{\C_j}^\star\|^2$, where $\C_i$ and $\C_j$ denote two neighboring clusters. Indeed, $\bw^\star_{\C_j}$ can be expressed as:
\begin{equation}
	\label{eq: task expression}
	\bw^\star_{\C_j}=\bw_o+R \left(
	\begin{split}
		\cos\Big(\theta_0+\frac{\theta}{Q}\big(j-\frac{1}{2}\big)\Big)\\
	\sin\Big(\theta_0+\frac{\theta}{Q}\big(j-\frac{1}{2}\big)\Big)
	\end{split}
	\right),\quad \forall j=1,\ldots,Q,
\end{equation}
where $\theta=\theta_1-\theta_0$. With the topology shown in Fig.\,\ref{fig:topology3}, we obtain:
\begin{equation}
\|\bw_{\C_i}^\star-\bw_{\C_j}^\star\|^2=R^2(2-2\cos(\theta/Q)).
\end{equation}
Figure~\ref{fig:MSD comparison} shows that cooperation among clusters improved the network MSD performance and endowed the network with robustness towards asynchronous events. We also observe that the performance of the standard diffusion LMS algorithm deteriorates when the level of similarity between tasks decreases. Figure~\ref{fig:estimation comparison} depicts the estimated arc when $R=R_0$ for the following algorithms in an asynchronous setting: noncooperative LMS obtained by setting $\bA(i)=\bP(i)=\bI_N$ for all $i$, standard diffusion LMS~\cite{Lopes2008diff}, and multitask diffusion LMS~\eqref{asynchronous_ATC}. In each case, the results were averaged over 150 Monte-Carlo runs and over 50 samples after convergence. The multitask diffusion algorithm outperformed the non cooperative LMS and the standard diffusion. The standard diffusion was not able to estimate the location of the target since it is a single task algorithm. It is shown in~\cite{chen2013pareto} that standard diffusion LMS converges to a Pareto optimal solution when it is applied to multitask problems.

\begin{figure*}
	\centering
	\includegraphics[scale=0.3]{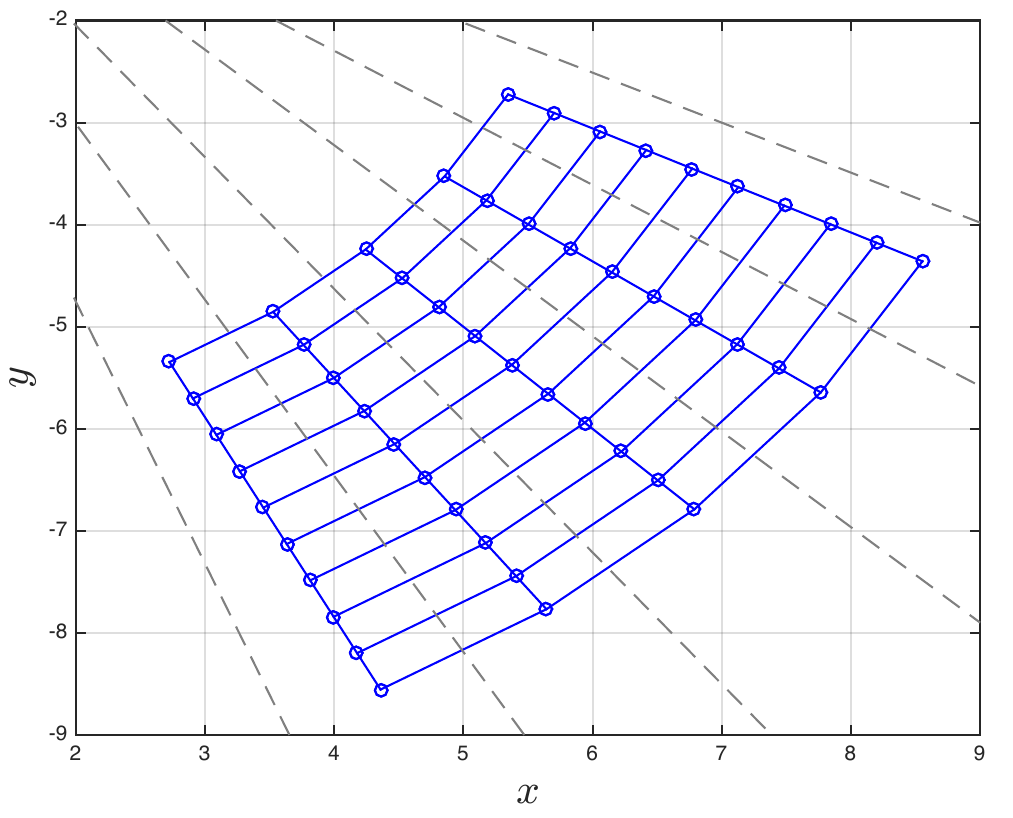}
	\includegraphics[scale=0.3]{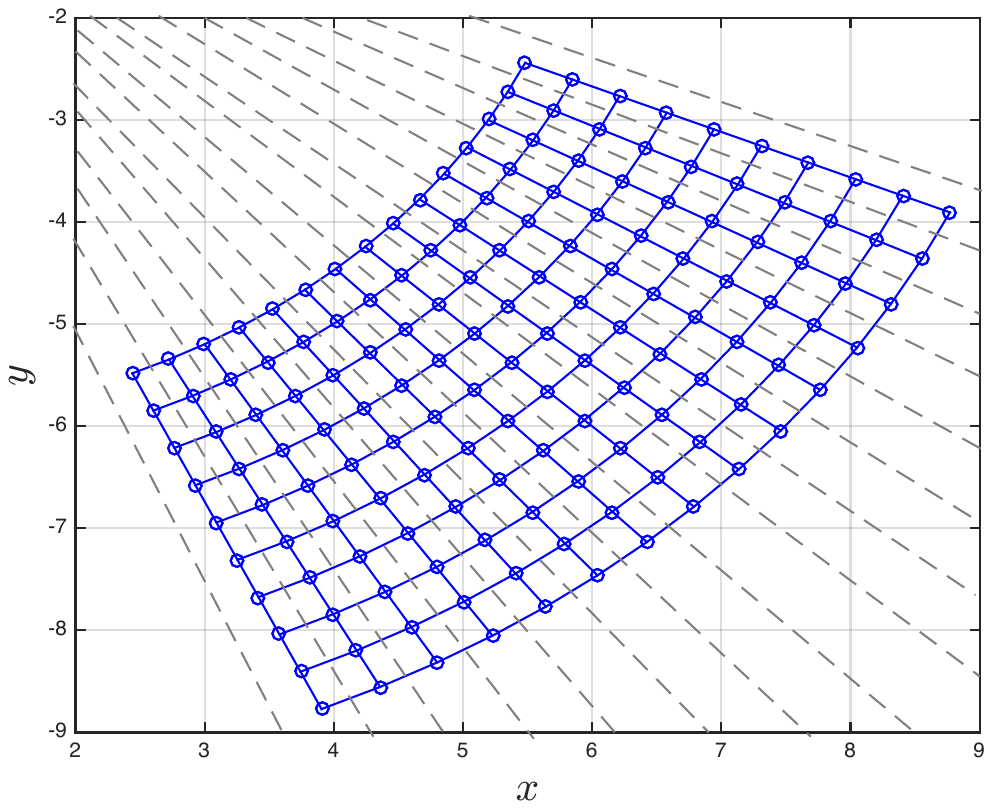}
	\caption{Network topology: circles for nodes, solid lines for links, and dashed lines for cluster boundaries. Left: network consisting of 5 clusters. Right: network consisting of 15 clusters.}
	\label{fig: network topologies 5 and 15 clusters}
\end{figure*}

\begin{figure*}
	\centering
	\includegraphics[scale=0.18]{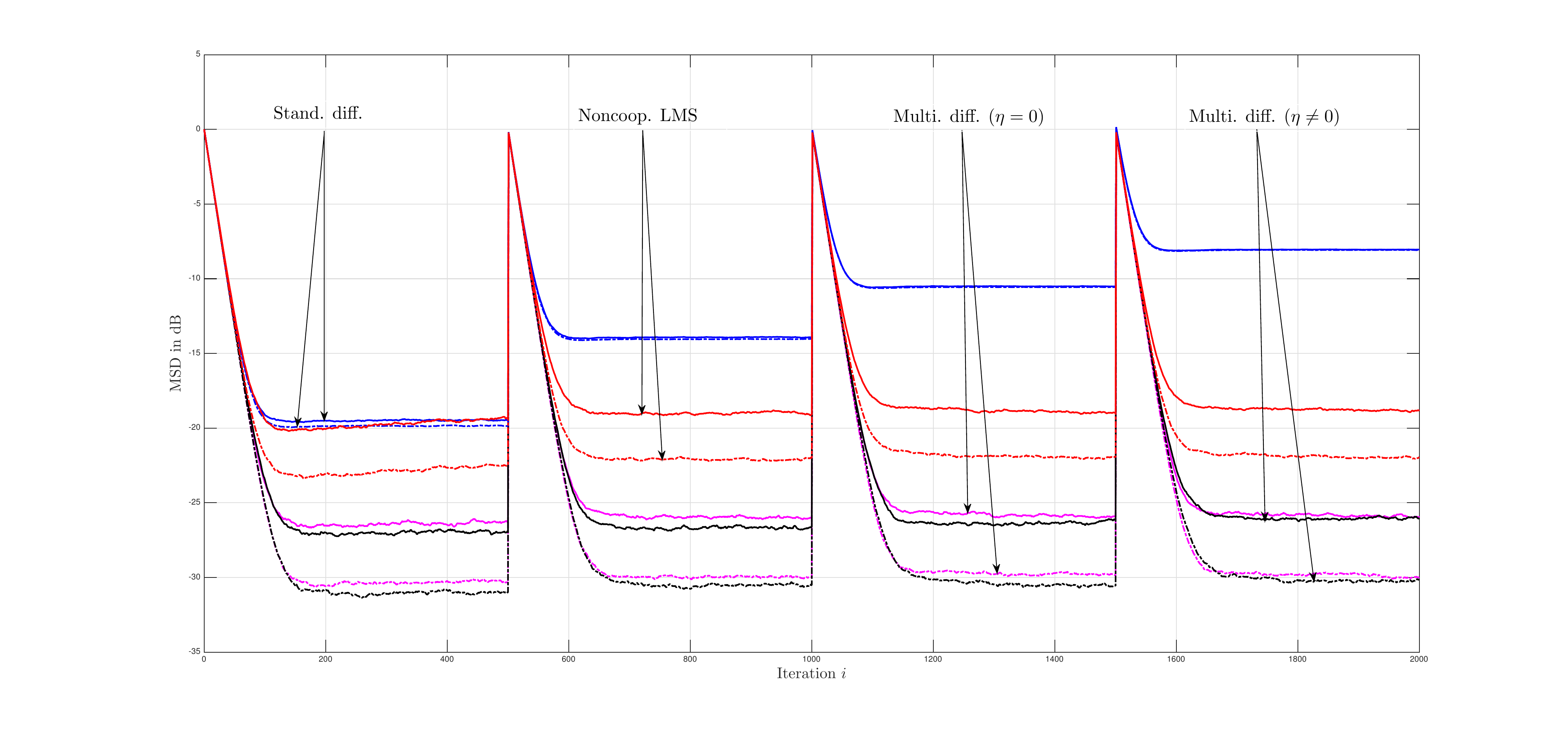}
	\includegraphics[scale=0.18]{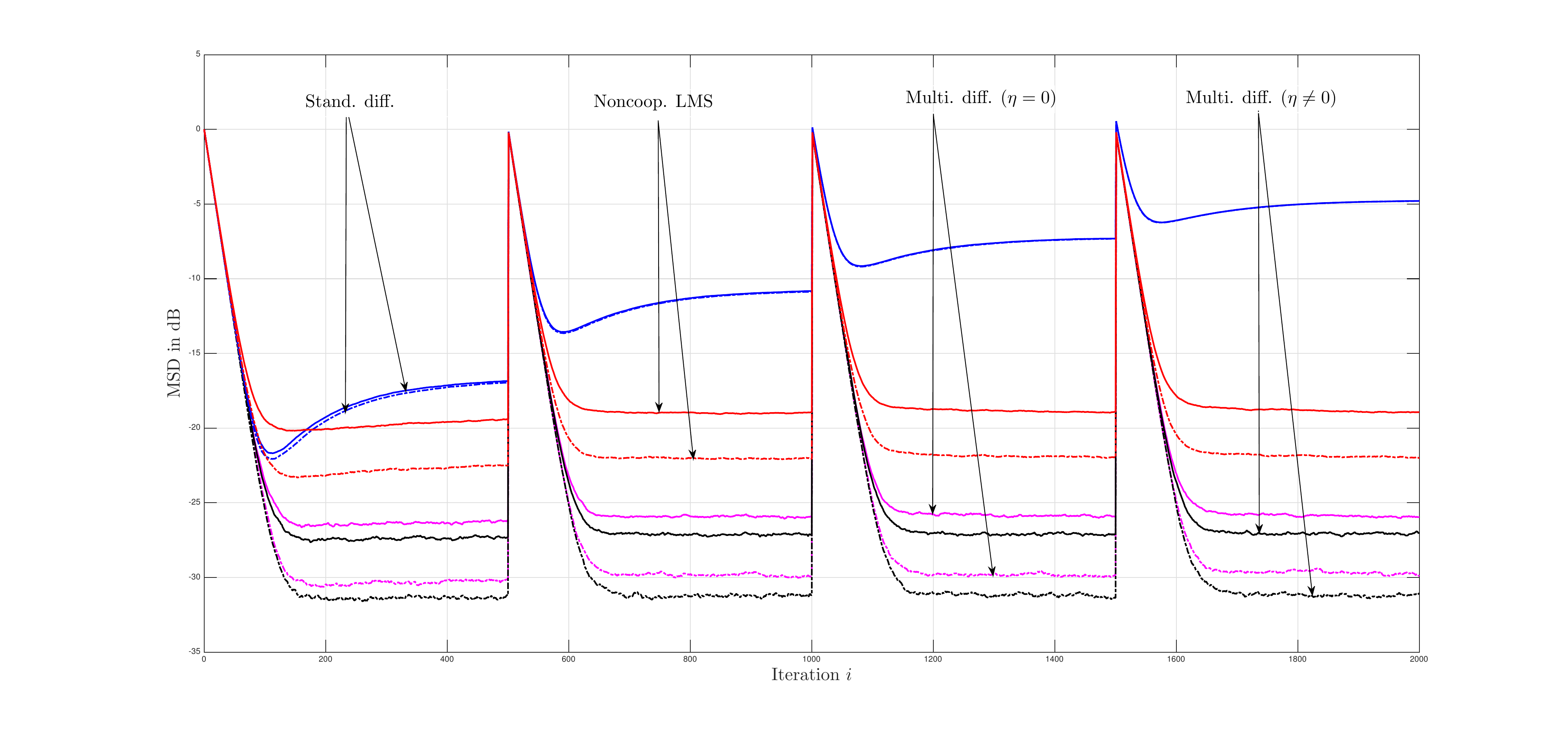}
	\caption{Network MSD learning curves in a non-stationary environment: comparison of the same algorithms considered in Figure~\ref{fig:MSD comparison}. The dotted lines are for synchronous networks and the solid lines are for asynchronous networks. Top: network consisting of $5$ clusters. Down: Network consisting of $15$ clusters.}
	\label{fig:MSD comparison 5 clusters and 15 clusters}
\end{figure*}

Finally, in order to show the effects of the number of clusters (or tasks) on the performance of the network, we considered 2 additional experimental setups. In the first one represented in Figure~\ref{fig: network topologies 5 and 15 clusters} (left), the number of tasks was set to $5$, that is, the arc $\mathcal{L}$ was decomposed into 5 sub-arcs. In the second one depicted in Figure~\ref{fig: network topologies 5 and 15 clusters} (right), the number of clusters was set to 15. Except for these changes, we considered the same experimental setup as before. Every $500$ time steps, the radius $R$ of the arc was modified as before in order to decrease the similarity level between tasks. The learning curves of the algorithms considered in Figure~\ref{fig:MSD comparison} are reported in~Figure~\ref{fig:MSD comparison 5 clusters and 15 clusters}. As expected, it can be observed that the larger the number of clusters is, the more efficient the collaboration between clusters becomes. The benefits of inter-cluster cooperation decreases when the number of clusters becomes small.

\section{Conclusion and perspectives}

In this paper, we considered multitask problems where networks are able to handle situations beyond the case where the nodes estimate a unique parameter vector over the network. We introduced a general model for asynchronous behavior with random step-sizes, combination coefficients, and co-regularization factors. We then carried out a convergence analysis of the asynchronous multitask algorithm in the mean and mean-square-error sense, and we derived conditions for convergence. Several open problems still have to be solved for specific applications. For instance, it would be interesting to investigate how nodes can autonomously adjust co-regularization factors between neighboring clusters in order to optimize the learning performance. It would also be advantageous to consider alternative co-regularizers in order to promote properties such as sparsity or block sparsity, and to analyze the convergence behavior of the resulting algorithms.


\begin{appendices}
\section{The Bernoulli model} 
\label{bernoulli model}
In this model, the step-sizes $\{\mu_k(i)\}$ are distributed as follows:
\begin{equation}
\label{eq: PD mu}
	\mu_k(i)=\left\lbrace
	\begin{array}{ll}
		\mu_k,	&\text{with probability } q_k	\\
		0, 		&\text{with probability } 1-q_k	\\
	\end{array}\right.
\end{equation} 
where $\mu_k$ is a fixed value. This probability distribution allows us to model random ``on-off'' behavior by each agent $k$ due to power saving strategies or random agent failures. We assume that the step-sizes $\mu_k(i)$ are spatially uncorrelated for different~$k$. At each iteration $i$, the mean of the step-size $\mu_k(i)$ is $\bar\mu_k=\mu_k q_k$, and the covariance between $\mu_k(i)$ and $\mu_\ell(i)$ is:
\begin{align}
	c_{\mu,k,\ell}&\triangleq\mathbb{E}\{(\mu_k(i)-\bar\mu_k)(\mu_\ell(i)-\bar\mu_\ell)\}\nonumber\\
	&=\left\lbrace\begin{array}{lr}
	\mu_k^2 q_k (1-q_k),\quad \text{if } \ell=k\\
	0,\quad\quad\quad\quad\quad\quad\text{otherwise.}\\
\end{array}
\right.
\end{align}
Furthermore, combination weights $\{a_{\ell k}(i)\}$ are distributed as follows:
\begin{equation}
\label{eq: PD a}
	a_{\ell k}(i)=\left\lbrace
	\begin{array}{lll}
		a_{\ell k}, 	& 	\text{with probability } p_{\ell k}\\
		0, 		&	\text{with probability } 1-p_{\ell k}\\
	\end{array}\right.
\end{equation} 
for any $\ell\in \N^-_{k}(i)\cap\C(k)$, where $0<a_{\ell k}<1$ a fixed coefficient. The coefficients $\{a_{\ell k}(i)\}$ are spatially uncorrelated for different $\ell$ and $k$. Node $k$ adjusts its own combination coefficient to ensure that the sum of its neighboring coefficients is equal to one as follows:
\begin{equation}
	a_{kk}(i)=1-\sum_{\ell\in \N^-_{k}(i)\cap \C(k)} a_{\ell k}(i)\,\geq 0.
\end{equation}
The probability distribution \eqref{eq: PD a} allows us to model a random ``on-off" status for links within clusters at time $i$ due to communication cost saving strategies or random link failures. With this model, we are giving the opportunity to each agent $k$ to randomly choose a subset of  neighbors that belong to its cluster to perform the combination step. At each iteration $i$, the mean of the coefficient $a_{\ell k}(i)$ is given by:
\begin{equation}
	\bar{a}_{\ell k}=
	\left\lbrace\begin{array}{ll}
	a_{\ell k}p_{\ell k},&								\text {if} \quad \ell\in \N^-_k\cap \C(k)\\
	1-\sum\limits_{\ell\in \N_k^-\cap \C(k)}a_{\ell k}p_{\ell k},&		\text{if}\quad \ell=k\\
	0,&											\text{otherwise}.
	\end{array}\right.
\end{equation}
and the covariance between $a_{\ell k}(i)$ and $a_{nm}(i)$ equals \cite{SayedPart1}:
\begin{align}
	&c_{a,\ell k,nm}=\mathbb{E}\{(a_{\ell k}(i)-\overline{a}_{\ell k})(a_{nm}(i)-\overline{a}_{nm})\}\nonumber\\
			&=
	\left\lbrace\begin{array}{ll}
	c_{a,\ell k,\ell k},&					\text{if }k=m,\ell=n, \ell \in \N^-_k\cap\C(k)\\
	-c_{a,\ell k,\ell k},&					\text{if }k=m=n, \ell \in \N^-_k\cap\C(k)\\
	-c_{a,nk,nk},&						\text{if }k=m=\ell, n \in \N^-_k\cap\C(k)\\
	\sum\limits_{j\in \N^-_k\cap\C(k)}c_{a,jk,jk},&	\text{if } k=m=\ell=n\\
	0,&								\text{otherwise.}
	\end{array}
\right.
\end{align}
where $c_{a,\ell k,\ell k}=a_{\ell k}^2p_{\ell k}(1-p_{\ell k})$.

Finally, the regularization factors $\{\rho_{k\ell}(i)\}$ are distributed as follows:
\begin{equation}
\label{eq: PD rho}
	\rho_{k\ell}(i)=\left\lbrace
	\begin{array}{ll}
		\rho_{k\ell},& 	\text{with probability } r_{k\ell}\\
		0,&			\text{with probability } 1-r_{k\ell}\\
	\end{array}\right.
\end{equation}
for any $\ell\in \N_{k}(i)\setminus\C(k)$, where $0<\rho_{k\ell}<1$ is a fixed regularization factor. The factors $\{\rho_{k\ell}(i)\}$ are spatially uncorrelated for $k\neq\ell$. At each iteration $i$, in order to get a right stochastic matrix $\bP(i)$, node $k$ adjusts its regularization factor as follows:
\begin{equation}
	\rho_{kk}(i)=1-\sum_{\ell\in \N_{k}(i)\setminus \C(k)} \rho_{k\ell}(i)\,\geq 0.
\end{equation}
The probability distribution \eqref{eq: PD rho} allows each agent $k$ to randomly select a subset of neighbors that do not belong to its cluster and introduce co-regularization in the estimation process. This behavior can also be interpreted as resulting from link random failures between neighboring clusters: at every time instant $i$, the communication link from agent $\ell$ to agent $k$ drops with probability $1-r_{k\ell}$. The mean of $\rho_{k\ell}(i)$ is given:
\begin{equation}
	\overline{\rho}_{k\ell }=\left\lbrace
	\begin{array}{ll}
		\rho_{k\ell}r_{k\ell},&									\text {if } \ell\in \N_k\setminus \C(k)\\
		1-\sum\limits_{\ell\in \N_k\setminus \C(k)}\rho_{k\ell}r_{k\ell},&		\text{if}\quad \ell=k\\
		0,&												\text{otherwise},
	\end{array}\right.
\end{equation} 
and the covariance between $\rho_{k\ell}(i)$ and $\rho_{mn}(i)$ is:
\begin{align}
	&c_{\rho,k\ell,mn}=\mathbb{E}\{(\rho_{k\ell}(i)-\overline{\rho}_{k\ell })(\rho_{mn}(i)-\overline{\rho}_{mn})\}\nonumber\\&=\left\lbrace
	\begin{array}{ll}
		c_{\rho,k\ell,k\ell},&						\text{if }k=m,\ell=n, \ell \in \N_k\setminus\C(k)\\
		-c_{\rho,k\ell,k\ell},&						\text{if }k=m=n, \ell \in \N_k\setminus\C(k)\\
		-c_{\rho,kn,kn},&						\text{if }k=m=\ell, n \in \N_k\setminus\C(k)\\
		\sum\limits_{j\in \N_k\setminus\C(k)}c_{\rho,kj,kj},&	\text{if } k=m=\ell=n\\
		0,	&								\text{otherwise}
	\end{array}
	\right.
\end{align}
where $c_{\rho,k\ell,k\ell}=\rho_{k\ell}^2 r_{k\ell}(1-r_{k\ell})$.
\section{Stability of $\cbp{F}$} 
\label{stability of F}

Recall from \eqref{eq:approxF} that
\begin{align}
	\cbp{F}\approx\cbp{A}_{\text{\,I}}^\top[&\bI_{(NL)^2}-\bI_{NL}\otimes_b\cbp{M}(\cbp{R}_x+\eta\cbp{Q})-\nonumber\\
	&\cbp{M}(\cbp{R}_x+\eta\cbp{Q})\otimes_b\bI_{NL}].
\end{align}
We now upper-bound the spectral radius of $\cbp{F}$ in order to derive a sufficient condition for mean-square stability of the algorithm. We can write:
\begin{align}
	\label{eq:upperF-1}
	\rho(\cbp{F})\leq\|\cbp{A}_{\text{\,I}}^\top\|_{b,\infty}\cdot\|&\bI_{(NL)^2}-\bI_{NL}\otimes_b\cbp{M}(\cbp{R}_x+\eta\cbp{Q})-\nonumber\\
	&\cbp{M}(\cbp{R}_x+\eta\cbp{Q})\otimes_b\bI_{NL}\|_{b,\infty}
\end{align}
Since the matrix $\cbp{A}_{\text{\,I}}$ is a block left-stochastic matrix, we know that $\|\cbp{A}_{\text{\,I}}^\top\|_{b,\infty}=1$. Using \eqref{eq: mean Q} and the triangular inequality, we have:
\begin{align}
	\label{eq:upperF-2}
	\rho(\cbp{F})	&\leq\|\bI_{(NL)^2}-\bI_{NL}\otimes_b\cbp{M}(\cbp{R}_x+\eta\bI_{NL})-\nonumber\\
					&\quad\quad\quad\quad\quad\cbp{M}(\cbp{R}_x+\eta\bI_{NL})\otimes_b\bI_{NL}\|_{b,\infty} \notag\\
				&+\eta\|\bI_{NL}\otimes_b\cbp{MP}\|_{b,\infty}
					+\eta\|\cbp{M}\cbp{P}\otimes_b\bI_{NL}\|_{b,\infty}.
\end{align}
Consider the second term on the RHS of \eqref{eq:upperF-2}. We know that
\begin{align}
		\bI_{NL}\otimes_b\cbp{MP}	
			&\overset{\eqref{property 3}}{=}(\bI_{NL}\otimes_b\cbp{M})(\bI_{NL}\otimes_b\cbp{P})\nonumber \\
			&\overset{\eqref{property 4}}{=}\Big((\bI_{N}\otimes\overline{\bM})\otimes\bI_{L^2}\Big)\Big((\bI_{N}\otimes\overline{\bP})\otimes\bI_{L^2}\Big).
\end{align}
Since $\Big((\bI_{N}\otimes\overline{\bP})\otimes\bI_{L^2}\Big)$ is a block right-stochastic matrix and $\Big((\bI_{N}\otimes\overline{\bM})\otimes\bI_{L^2}\Big)$ is an $N^2\times N^2$ block diagonal matrix with each block of the form $\bar\mu_k\bI_{L^2}$ ($k=1,\ldots,N$), we obtain:
\begin{align}
	\label{eq:stabF-t2}
		&\|\bI_{NL}\otimes_b\cbp{MP}\|_{b,\infty}\nonumber\\	
			&\leq	\|(\bI_{N}\otimes\overline{\bM})\otimes\bI_{L^2}\|_{b,\infty}\cdot\|(\bI_{N}\otimes\overline{\bP})\otimes\bI_{L^2}\|_{b,\infty}\nonumber	\\
			&=\max_{1\leq k\leq N} \overline {\mu}_k
\end{align}
Following the same steps for the third term on the RHS of \eqref{eq:upperF-2}, we have:
\begin{equation}
	\label{eq:stabF-t3}
	\|\cbp{MP}\otimes_b\bI_{NL}\|_{b,\infty}\leq\max_{1\leq k\leq N} \overline {\mu}_k.
\end{equation}
The matrix $\big[\bI_{(NL)^2}-\bI_{NL}\otimes_b\cbp{M}(\cbp{R}_x+\eta\bI_{NL})-\cbp{M}(\cbp{R}_x+\eta\bI_{NL})\otimes_b\bI_{NL}\big]$ in the first term on the RHS of \eqref{eq:upperF-2} is an $N^2\times N^2$ block diagonal matrix. The $m$-th block on the diagonal (where $m=(\ell-1)N+k$ for $k,\ell=1,\ldots,N$) is of size $L^2\times L^2$, symmetric, and has the following form:
\begin{align}
	\label{eq:m-th block entry}
	&\bI_{L^2}-\bI_L\otimes\bar\mu_k(\boldsymbol {R}_{x,k}+\eta\bI_L)-\bar\mu_\ell(\boldsymbol {R}_{x,\ell}+\eta\bI_L)\otimes\bI_L\nonumber\\
	&\hspace{-1.25mm}=\hspace{-0.75mm}(\hspace{-0.25mm}-\bar\mu_\ell\boldsymbol {R}_{x,\ell}-\eta\bar\mu_\ell\bI_L)\hspace{-0.25mm}\otimes
	\bI_L+\bI_L\hspace{-0.25mm}\otimes(\bI_L-\bar\mu_k\boldsymbol {R}_{x,k}-\eta\bar\mu_k\bI_L)
\end{align}
Before proceeding, let us recall the Kronecker sum operator, denoted by $\oplus$. If $\bA$ and $\bB$ are two matrices of dimension $L\times L$ each, then
\begin{equation}
\bA \oplus \bB\triangleq\bA \otimes \bI_L+ \bI_L \otimes \bB.
\end{equation}
Let $\lambda_k\{\cdot\}$ denote the $k$-th eigenvalue of its matrix argument. Then, the eigenvalues of $\bA\oplus\bB$ are of the form $\lambda_i\{\bA\}+\lambda_j\{\bB\}$ for $i$, $j=1,\ldots,L$ \cite{bernstein2005matrix}. Note that the RHS of equation \eqref{eq:m-th block entry} can be written as 
\begin{equation}
(-\bar\mu_\ell\boldsymbol {R}_{x,\ell}-\eta\bar\mu_\ell\bI_L)\oplus(\bI_L-\bar\mu_k\boldsymbol {R}_{x,k}-\eta\bar\mu_k\bI_L)
\end{equation}
and its eigenvalues are therefore of the form:
\begin{equation}
	\label{eq:vapkroneckersum}
	1-\eta\bar\mu_k-\bar\mu_k\lambda_j\{\bR_{x,k}\}
	-\eta\bar\mu_\ell-\bar\mu_\ell\lambda_i\{\boldsymbol {R}_{x,\ell}\}
\end{equation}
for $i$, $j=1,\ldots,L$ and $k$, $\ell=1,\ldots,N$. In order to simplify the mean-square stability condition, we assume that the first order moment of the step-sizes is the same for all nodes. Using the fact that the block maximum norm of a block diagonal Hermitian matrix is equal to the largest spectral radius of its block entries \cite{Sayed2013intr}, we get:
\begin{align}
	\label{eq:stabF-t1}
	&\|\bI_{(NL)^2}-\bI_{NL}\otimes_b\cbp{M}(\cbp{R}_x+\eta\bI_{NL})-\nonumber\\
	&\quad\quad\quad\quad\quad\quad\cbp{M}(\cbp{R}_x+\eta\bI_{NL})\otimes_b\bI_{NL}\|_{b,\infty}\nonumber\\
	&=\max_{1\leq k,\ell \leq N}\Big(\max_{1\leq i,j\leq L}|1-2\eta\bar\mu-\bar\mu(\lambda_j\{\bR_{x,k}\}+\lambda_i\{\boldsymbol {R}_{x,\ell}\})|\,\Big)\nonumber\\
	&=\max_{1\leq k,\ell \leq N}\Big(\max_{1\leq i,j\leq L}\{1-2\eta\bar\mu-\bar\mu(\lambda_j\{\bR_{x,k}\}+\lambda_i\{\boldsymbol {R}_{x,\ell}\}),\nonumber\\ 
	&\quad\quad\quad\quad\quad\quad\quad\quad-1+2\eta\bar\mu+\bar\mu(\lambda_j\{\bR_{x,k}\}+\lambda_i\{\boldsymbol {R}_{x,\ell}\})\}\,\Big)\nonumber\\
	&=\max\,\{1-2\eta\bar\mu-\bar\mu\min_{k,\ell}\,(\lambda_{\min}\{\bR_{x,k}\}+\lambda_{\min}\{\boldsymbol {R}_{x,\ell}\}),\nonumber\\
	&\quad\quad\quad-1+2\eta\bar\mu+\bar\mu\max_{k,\ell}\,(\lambda_{\max}\{\bR_{x,k}\}+\lambda_{\max}\{\boldsymbol {R}_{x,\ell}\})\}.
\end{align}
The minimum (identically the maximum) on $k$ and $\ell$ that appears in the last equality of \eqref{eq:stabF-t1} is reached for $k=\ell$. Thus, a sufficient condition for mean-square stability is given by:
\begin{eqnarray}
	\max_{1\leq k\leq N}(\max_{1\leq i\leq L}|1-2\eta\bar\mu-2\bar\mu\lambda_i(\bR_{x,k})|
		+2\eta  {\bar\mu}) <1,
\end{eqnarray}
which is verified if the first order moment of the step-sizes satisfies:
\begin{equation}
0<\bar\mu< \frac{1}{2\eta+\max_{1\leq k \leq N}\rho(\bR_{x,k})}.
\end{equation}

\end{appendices}

\bibliographystyle{IEEEbib}
\bibliography{asynchronous}

\balance
\begin{IEEEbiography}[{\includegraphics[width=1in,height=1.25in,clip,keepaspectratio]{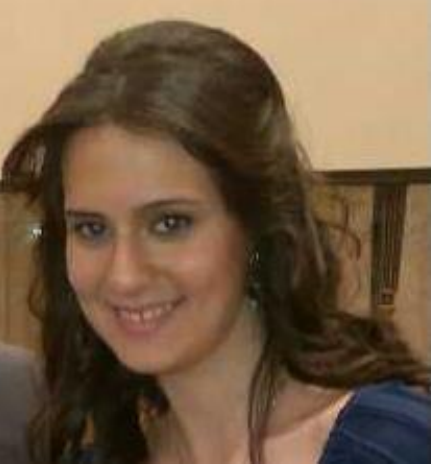}}
]{Roula Nassif}

was born in Beirut, Lebanon in February 1991. She received the bachelor's degree in Electrical Engineering from the Lebanese University, Lebanon, in 2013. She received the M.S. degrees in Industrial Control and Intelligent Systems for Transport from the Lebanese University, Lebanon, and from Compi\`egne University of Technology, France, in 2013. Since October 2013 she is a Ph.D. student at the Lagrange Laboratory (University of Nice Sophia Antipolis, CNRS, Observatoire de la C\^ote d'Azur). Her research activity is focused on distributed optimization over multitask networks. 

\end{IEEEbiography}

\begin{IEEEbiography}[{\includegraphics[width=1in,height=1.25in,clip,keepaspectratio]{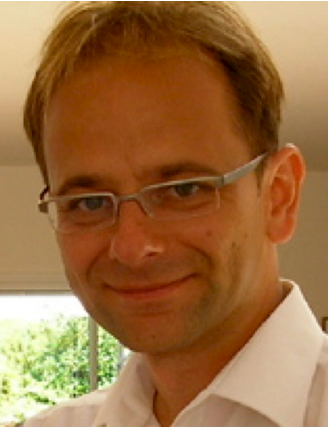}}]{C{\'e}dric Richard}

(S'98--M'01--SM'07)  received the Dipl.-Ing. and the M.S. degrees in 1994, and the Ph.D. degree in 1998, from Compi\`egne University of Technology, France, all in electrical and computer engineering. From 1999 to 2003, he was an Associate Professor at Troyes University of Technology, France, and a Full Professor from 2003 to 2009. Since 2009, he is a Full Professor at the University of Nice Sophia Antipolis, France. He was a junior member of the Institut Universitaire de France in 2010-2015.

His current research interests include statistical signal processing and machine learning. C\'edric Richard is the author of over 230 papers. He was the General Co-Chair of the IEEE SSP Workshop that was held in Nice, France, in 2011. He was the Technical Co-Chair of EUSIPCO 2015 that was held in Nice, France, and of the IEEE CAMSAP Workshop 2015 that was held in Cancun, Mexico. He serves as a Senior Area Editor of the IEEE Transactions on Signal Processing and as an Associate Editor of the IEEE Transactions on Signal and Information Processing over Networks since 2015. He is also an Associate Editor of Signal Processing Elsevier since 2009. C\'edric Richard is member of the Machine Learning for Signal Processing (MLSP TC) Technical Committee, and served as member of the Signal Processing Theory and Methods (SPTM TC) Technical Committee in 2009-2014.

\end{IEEEbiography}
\balance

\begin{IEEEbiography}[{\includegraphics[width=1in,height=1.25in,clip,keepaspectratio]{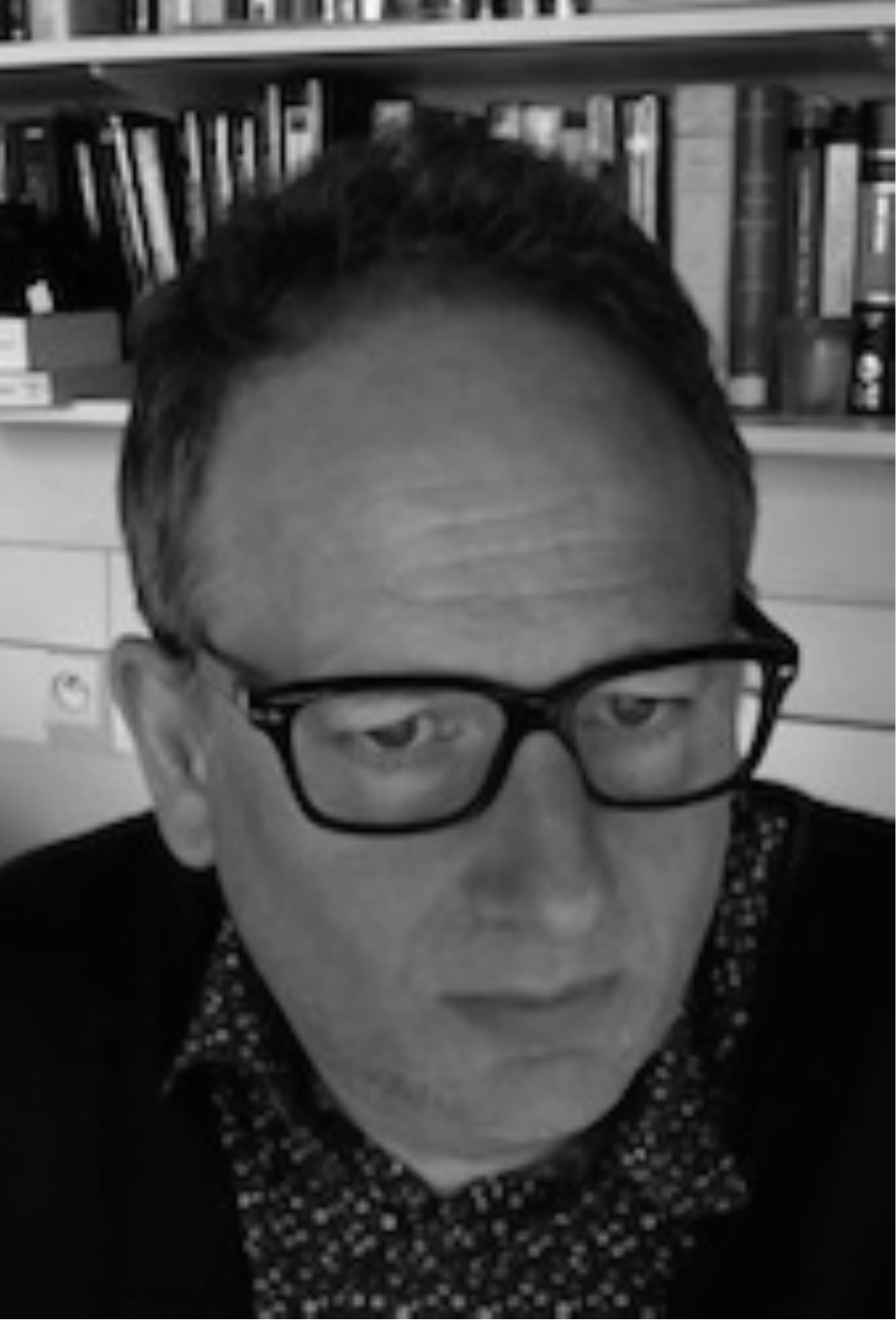}}
]{Andr{\'e} Ferrari}

(SM'91-M'93) received the Ing\'enieur degree from \'Ecole Centrale de Lyon, Lyon, France, in 1988 and the M.Sc. and Ph.D. degrees from the University of Nice Sophia Antipolis (UNS), France, in 1989 and 1992, respectively, all in electrical and computer engineering. He is currently a Professor at UNS.

He is currently a Professor at UNS. He is a member of the Joseph-Louis Lagrange Laboratory (CNRS, OCA), where his research activity is centered around statistical signal processing and modeling, with a particular interest in applications to astrophysics.

\end{IEEEbiography}

\begin{IEEEbiography}[{\includegraphics[width=1in,height=1.25in,clip,keepaspectratio]{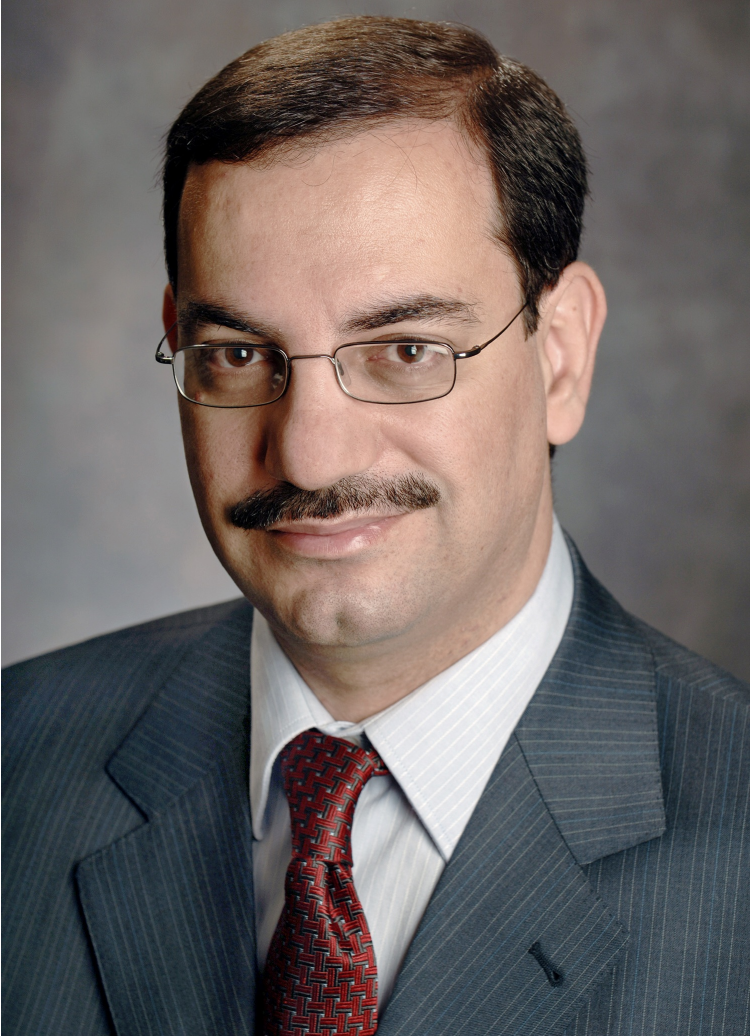}}
]{Ali H. Sayed}

(S'90-M'92-SM'99-F'01) is professor and former chairman of electrical engineering at the University of California, Los Angeles, USA, where he directs the UCLA Adaptive Systems Laboratory. An author of more than 460 scholarly publications and six books, his research involves several areas including adaptation and learning, statistical signal processing, distributed processing, network science, and biologically inspired designs. Dr. Sayed has received several awards including the 2015 Education Award from the IEEE Signal Processing Society, the 2014 Athanasios Papoulis Award from the European Association for Signal Processing, the 2013 Meritorious Service Award, and the 2012 Technical Achievement Award from the IEEE Signal Processing Society. Also, the 2005 Terman Award from the American Society for Engineering Education, the 2003 Kuwait Prize, and the 1996 IEEE Donald G. Fink Prize. He served as Distinguished Lecturer for the IEEE Signal Processing Society in 2005 and as Editor-in Chief of the IEEE TRANSACTIONS ON SIGNAL PROCESSING (2003-2005). His articles received several Best Paper Awards from the IEEE Signal Processing Society (2002, 2005, 2012, 2014). He is a Fellow of the American Association for the Advancement of Science (AAAS). He is recognized as a Highly Cited Researcher by Thomson Reuters.
\end{IEEEbiography}

\end{document}